\documentclass{article}

\PassOptionsToPackage{numbers, compress}{natbib}

\usepackage[preprint]{neurips_2026}

\usepackage[utf8]{inputenc} 
\usepackage[T1]{fontenc}    
\usepackage{microtype}
\usepackage{graphicx}
\usepackage{caption}
\usepackage{subcaption}
\usepackage{booktabs} 
\usepackage{multirow}
\usepackage{hyperref}
\usepackage{url}            
\usepackage{amsfonts}       
\usepackage{nicefrac}       
\usepackage{xcolor}         
\usepackage{algorithm}
\usepackage[noend]{algorithmic}
\usepackage{wrapfig}

\usepackage{amsmath}
\usepackage{amssymb}
\usepackage{mathtools}
\usepackage{amsthm}
\usepackage{enumitem}
\usepackage{pifont}
\newcommand{\cmark}{\ding{51}}%
\newcommand{\xmark}{\ding{55}}%

\usepackage[normalem]{ulem}
\useunder{\uline}{\ul}{}

\DeclareMathOperator{\sigmoid}{sigmoid}

\usepackage[capitalize,noabbrev]{cleveref}

\theoremstyle{plain}

\theoremstyle{definition}

\theoremstyle{remark}

\usepackage[textsize=tiny]{todonotes}
\usepackage{xspace}
\usepackage{tabularx}
\newcommand{\ourmodel}{A-CODE\xspace}

\title{A-CODE: Fully Atomic Protein Co-Design with Unified Multimodal Diffusion}

\author{%
  Chaoran Cheng$^1$\thanks{Equal contribution.}\:\:\thanks{Work done during internship at ByteDance Seed. $^\ddagger$Work done during employment at ByteDance Seed.}\quad
  Jiaqi Guan$^2$\footnotemark[1]\quad
  Milong Ren$^2$\footnotemark[2]\quad
  Chengyue Gong$^2$\quad
  Cong Liu$^2$\footnotemark[2]\\
  {\bf
  Xinshi Chen$^2$\footnotemark[3]\quad
  Ge Liu$^1$\quad
  Wenzhi Xiao$^2$\footnotemark[3]
  }\\
  $^1$University of Illinois Urbana-Champaign \quad
  $^2$ByteDance Seed
}

\begin{document}

\maketitle
{
\renewcommand{\thefootnote}{\fnsymbol{footnote}}
\footnotetext[4]{Corresponding to \texttt{chaoran7@illinois.edu}.}
}

\begin{abstract}
We present A-CODE, a fully atomic unified one-stage protein co-design model that simultaneously refines discrete atom types and continuous atom coordinates. Unlike predominant two-stage methods that cascade structure design with amino acid-level sequence design, our approach is fully atomic within a unified multimodal diffusion framework, in which residue identities are inferred solely from atom-level predictions.
Built upon the powerful all-atom architecture, A-CODE achieves superior designability for unconditional protein generation, outperforming all existing one-stage and two-stage design models. For binder design, A-CODE rivals and even outperforms existing state-of-the-art two-stage design models and, compared with the existing one-stage co-design model, achieves a drastic tenfold improvement in success rate on hard tasks.
The inherent flexibility of our atomic formulation enables, for the first time, seamless adaptation to non-canonical amino acid (ncAA) modeling.
Our fully atomic framework establishes a new, versatile foundation for all-atom generative modeling that can be naturally extended to complex biomolecular systems.
\end{abstract}

\section{Introduction}

Proteins are the fundamental molecular machines of life, orchestrating nearly every biological process through their intricate three-dimensional structures and dynamic interactions. The ability to design novel proteins with properties beyond those found in nature holds transformative potential for science and medicine, promising breakthroughs such as highly specific therapeutics and novel biocatalysts. However, as a protein's function is strictly dictated by its sequence of amino acids, which folds into a unique geometric configuration, navigating the vast combinatorial space of possible sequences remains a formidable computational challenge.

The landscape of computational biology was fundamentally altered by the emergence of AlphaFold 2 (AF2)~\cite{jumper2021highly}, which demonstrated that deep neural networks could learn the intricate mapping from sequence to structure with near-experimental accuracy and was further extended to possess the unified all-atom predictive power in AlphaFold 3~\cite{abramson2024accurate} (AF3) more recently. This predictive breakthrough has catalyzed a new wave of generative approaches, with attempts to adapt such architectures originally designed for structure prediction into tools for protein design. Despite this progress, the prevailing paradigm for \textit{de novo} design remains largely bifurcated.

\begin{figure}[ht]
\centering
\includegraphics[width=\linewidth]{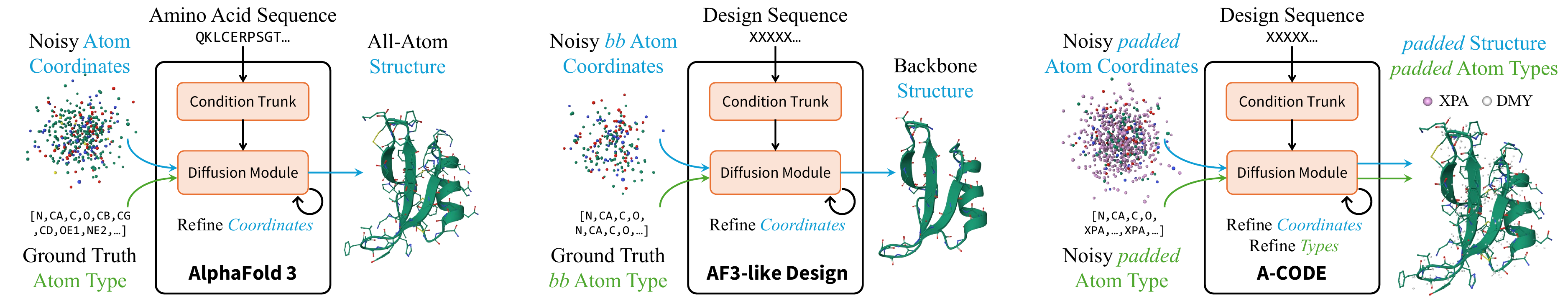}
\caption{Comparison of the model frameworks of AlphaFold 3 (\textbf{left}), AlphaFold 3-like design model (\textbf{middle}), and our \ourmodel (\textbf{right}) with atomic-scale co-design. Note how the initial noisy input features differ across the three models.}
\label{fig:model}
\end{figure}

Current state-of-the-art methods predominantly operate as \emph{two-stage} cascades: they first generate a geometric backbone structure—often via diffusion or flow matching and subsequently employ a separate inverse folding model to condition on that fixed backbone and predict a compatible amino acid sequence. For example, PXDesign~\cite{team2025pxdesign} extends Protenix~\cite{bytedance2025protenix}, an open-source reimplementation of AF3, for backbone-only structure design. 
More recent works like RFDiffusion3 ~\cite{butcher2025novo} and BoltzGen~\cite{stark2025boltzgen} that attempt to leverage the all-atom power of AF3-like models still suffer from such sequential decoupling that prevents the model from capturing the full joint distribution of sequence and structure. The latent-based generative models, such as La-Proteina~\cite{geffner2025proteina} and PLAID~\cite{lu2024generating}, offer an alternative, more flexible approach but remain limited by the accuracy of their structure VAEs at recovering all-atom structures.
On the other hand, a true \emph{one-stage} framework capable of unified co-design, in which geometry and biochemical semantics are generated simultaneously in an end-to-end manner, remains a significant open challenge in the field. 
While recent attempts such as MultiFlow~\cite{campbell2024generative} and Protpardelle~\cite{chu2024all} have been made, these models are generally less favorable for co-designability, as coordinating the structure and sequence modalities can introduce subtle trade-offs.

\begin{wraptable}{r}{.65\textwidth}
\vspace{-.8em}
\centering
\caption{Comparisons between existing protein co-design models.}\label{tab:models}
\vspace{-0.5em}
\resizebox{\linewidth}{!}{
\small
\begin{tabular}{@{}lccc@{}}
\toprule
Model & All-Atom & One-Stage Co-Design & ncAA Modeling \\ \midrule
MultiFlow [\citenum{campbell2024generative}] & \xmark & \cmark & \xmark \\
Protpardelle [\citenum{chu2024all}] & \cmark & \cmark & \xmark \\
ProteinGenerator  [\citenum{lisanza2025multistate}] & \cmark & \cmark & \xmark \\
Pallatom [\citenum{qu2024p}] & \cmark & \xmark & \xmark \\
La-Proteina [\citenum{geffner2025proteina}] & \cmark & \xmark & \xmark \\
RFDiffusion3 [\citenum{butcher2025novo}] & \cmark & \xmark & \xmark \\
PXDesign [\citenum{team2025pxdesign}] & \xmark & \xmark & \xmark \\ \midrule
\textbf{\ourmodel (Ours)} & \cmark & \cmark & \cmark \\ \bottomrule
\end{tabular}
}
\end{wraptable}
To address the challenges in protein structure–sequence co-design, we propose \emph{\ourmodel} (\textbf{A}tomic \textbf{CO}-\textbf{DE}sign), a unified one-stage framework that operates directly at the atomic scale. \ourmodel combines mask diffusion for discrete atom types with standard diffusion for continuous atom coordinates, forming a unified multimodal diffusion process over atomic representations. During sampling, the atom types and coordinates are denoised simultaneously at each iterative diffusion step, enabling finer-grained interactions between modalities. 
Furthermore, unlike existing co-design approaches that rely on residue-level type information, \ourmodel infers residue identities purely from atom-level predictions. This atom-centric formulation removes hard constraints on residue types in existing work and naturally extends to modeling non-canonical amino acids (ncAAs) without architectural modifications. 

Through extensive experiments on unconditional protein generation and conditional binder design, together with an in-depth analysis of ncAA modeling behavior, we demonstrate that \ourmodel achieves strong empirical performance and provides a flexible, unified foundation for all-atom generative modeling of diverse biomolecules. 
For unconditional protein design, \ourmodel establishes a new state-of-the-art designability and codesignability performance, outperforming the second-best La-Proteina by a significant 6-9\% increase and achieving a 2.5-12x inference-time speedup.
For the binder design task, \ourmodel rivals and even outperforms existing two-stage models for the first time, achieving a tenfold improvement in success rate on hard tasks.
Additionally, for the first time, we present quantitative results for ncAA modeling, in which \ourmodel demonstrates the ability to generate structurally plausible ncAA. In this way, our fully atomic design paves the way for a new paradigm to inspire future exploration of a unified all-atom design framework.

We summarize the main features of our \ourmodel with existing representative protein co-design models in Table~\ref{tab:models}, and we highlight our contributions in the following aspects:
\vspace{-0.5em}
\begin{itemize}[noitemsep,leftmargin=*]
    \item We present \ourmodel, a fully atomic, one-stage, unified protein co-design framework that combines discrete diffusion for atom types with continuous diffusion for atom coordinates.
    \item \textit{In silico} evaluations on the unconditional protein generation demonstrate \ourmodel's superior generation quality, with a significant 6-9\% increase in designability and codesignability and a 2.5-12x inference-time speedup compared to the second-best La-Proteina model.
    \item \textit{In silico} evaluations on the binder design task further validate \ourmodel's performance, rivaling and even outperforming existing state-of-the-art two-stage models. Compared to existing one-stage co-design models, a tenfold improvement in the success rates on hard tasks is observed.
    \item By predicting atom types without any hard constraint, our framework naturally extends to ncAA modeling, which existing models fail due to their hard-coded constraints.
\end{itemize}

\section{Related Work}
Recent advances in the generative framework, such as diffusion~\cite{ho2020denoising,song2020score,karras2022elucidating} and flow matching~\cite{lipman2022flow,chen2023flow}, have revolutionized generative modeling in all domains, including protein co-design, where each amino acid's position and type are learned. Noticeably, there are also existing works on extending such generative frameworks for the discrete modality~\cite{austin2021structured,cheng2024categorical,sahoo2024simple}.
While the AlphaFold 2~\cite{jumper2021highly} model still relies on multiple rounds of structural recycling, the all-atom AlphaFold 3~\cite{abramson2024accurate} already adopts the diffusion framework. Many protein co-design models are built on the expressive power of AF3-like models, which we detail below. Additional works are discussed in Appendix~\ref{suppl:related}.

As briefly discussed in the previous text, we categorize co-design models into \emph{one-stage} and \emph{two-stage} based on their approach to discrete sequence modality. The two-stage representatives include PXDesign~\cite{team2025pxdesign} and ODesign~\cite{zhang2025odesign}, which adopt a backbone-only design mode, whereas RFDiffusion-3~\cite{butcher2025novo} and BoltzGen~\cite{stark2025boltzgen} have all-atom design capability. These three models are built on AF3-like structures, leveraging its strong structure prediction performance, while other models develop their own architectures. La-Proteina~\cite{geffner2025proteina} and PLAID~\cite{lu2024generating} leverage latent flow matching and diffusion for determining the residue type and the all-atom structures based on the latents. Pallatom~\cite{qu2024p} relies on an alternative sequence decoder when maintaining the all-atom information. While existing all-atom design models predominantly rely on the 14-atom representation, APM~\cite{chu2024all} employs a side-chain structure module to model torsion angles.

In contrast, the one-stage co-design models remain relatively underexplored. MultiFlow~\cite{campbell2024generative} and ProteinGenerator~\cite{lisanza2025multistate} combine continuous flow matching for the backbone atom positions with discrete flow matching for amino acid types and update both modalities simultaneously during flow sampling. combined Protpardelle~\cite{chu2024all,lu2025conditional} relies on an exotic 73-atom presentation which combines all sidechain atoms from the 20 standard amino acids and selectively updates the corresponding one selected by the sequence predictor in each denoising step.
We also note that, in both two-stage and one-stage approaches, exploration of ncAA modeling remains extremely limited, often treating them as ligands and discarding residual-level information.

\section{\ourmodel: Fully Atomic Co-Design}

\begin{figure}
\centering
\includegraphics[width=\linewidth]{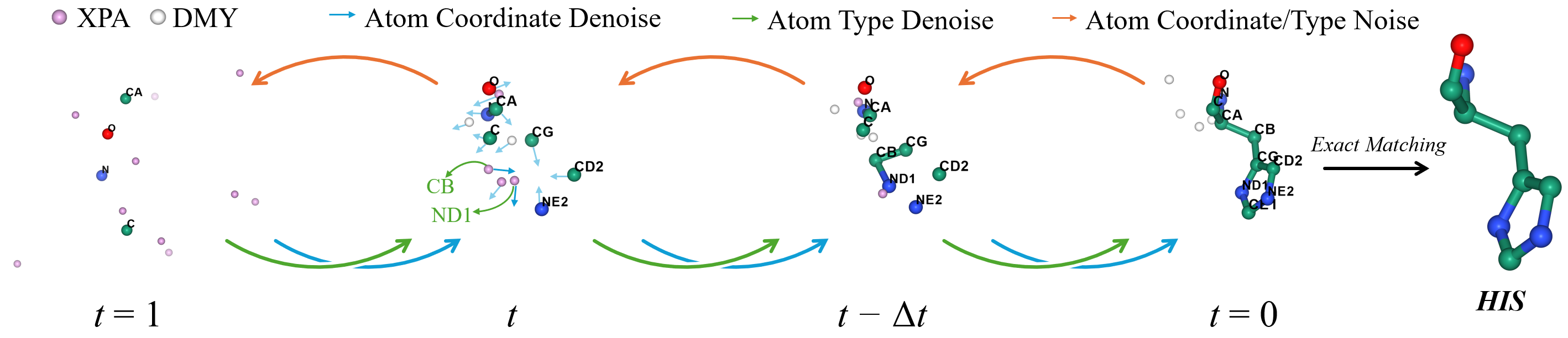}
\caption{Fully atomic sampling process with \ourmodel. Random noisy coordinates are sampled with 10 \texttt{XPA} tokens as the masks for the initial inputs. During each iterative sampling step, both atom types and coordinates are denoised simultaneously, thereby gradually generating a clean prediction.}
\label{fig:sampling}
\end{figure}

\subsection{Preliminary}
A protein can be described by its 3D structure information (atom coordinates) $X=\{\mathbf{x}_i\}_{i=1}^{n_\text{atom}},\mathbf{x}_i\in\mathbb{R}^3$ and amino acid sequence information (residue types) $S=\{c_i\}_{i=1}^{n_\text{res}},c_i\in\{\texttt{ALA},\texttt{ARG},\dots,\texttt{VAL}\}$. Each of the 20 canonical amino acids is associated with a unique list of atom types (names), which together build the discrete atom type information $A=\{a_i\}_{i=1}^{n_\text{atom}}$ that can be identified from $S$ or vice versa (See Appendix~\ref{suppl:data}, Table~\ref{tab:aa}).
Recent structure-prediction models, such as AF3, adopt a generative perspective that relies on diffusion to refine their atomic coordinates, thereby modeling $p(X|S)$. As a folding model, AF3's condition trunk first processes the input amino acid sequence $S$ together with other optional features, such as MSA and templates, to produce a conditioning representation $C$ for the diffusion module, which then takes in the noisy atom coordinates $\mathbf{x}$ and the residue-level representations $\mathbf{s}$ to predict clean coordinates:
\begin{equation}
    \hat{X_0}\gets\text{DenoiseNet}(X_t, S, \sigma_t, C),
\end{equation}
where $\sigma_t$ is the noise schedule.
The outline of the AF3 framework is visualized in Figure~\ref{fig:model}.
During sampling, the coordinates are updated according to the EDM scheduler~\cite{karras2022elucidating} using the clean data prediction, with the noise scale decreasing exponentially to produce gradually refined coordinates.

The protein design task requires modeling of joint distribution over structure and sequence modalities of $p(X,S)$. Existing work, such as PXDesign~\cite{team2025pxdesign}, predominantly adopts a two-stage approach and decomposes the joint distribution as $p(S|X)p(X)$. In this way, these approaches first rely on a structure-design model to generate $X$ from $p(X)$, and then leverage an inverse-folding model, such as ProteinMPNN (PMPNN)~\cite{dauparas2022robust}, to generate the sequence from $p(S|X)$ given the structure. 
Unlike PXDesign, our \ourmodel framework directly models the joint distribution $p(X,S)$, effectively capturing the all-atom structure and atom-type modalities simultaneously. We highlight the difference of \ourmodel to AF3 and AF3-derived design models in Figure~\ref{fig:model}, where the former uniquely refined the atom types together with atom coordinates during each iterative step. We now explain the design details of our unified atomic diffusion module.

\subsection{Unified Atomic Diffusion}

The core challenge of fully atomic protein co-design lies in jointly modeling structure and biochemical identity across heterogeneous, variable atomic compositions. Amino acids differ significantly in side-chain size and topology, which makes it non-trivial to define a unified representation. To address this, \ourmodel adopts the \emph{atom14} representation used in AF2, where each (canonical) amino acid is represented with at most 14 heavy atom coordinates. 
Unlike prior structure-only design models where \emph{dummy atoms} act purely as coordinate placeholders, in \ourmodel they are explicitly represented by a dedicated atom type (\texttt{DMY}) and are treated on equal footing with other atom types during diffusion. As a result, dummy atoms participate directly in the generative process and can also be predicted. This design allows residue identities to emerge naturally from atom-level predictions, thereby tightly coupling atomic structure generation and sequence inference within a single generative process.

Within this formulation, \ourmodel predicts atom names rather than coarse element or residue types. Atom names encode both chemical and positional semantics (e.g., CG, CD, ND1), which are essential for unambiguous residue identification. The four
backbone atoms (N, CA, C, O) are always present and therefore fixed during training and sampling, while the remaining side-chain positions are treated as designable and initialized with masked tokens  (\texttt{XPA0--XPA9}). Importantly, residue-level supervision for designable positions is entirely removed to prevent information leakage; residue types are inferred only after diffusion by matching the predicted atom-name sets. Together with the standard diffusion on the atom coordinates, the model operates purely in atom space, without directly optimizing or predicting residue labels. This atom-centric formulation removes hard constraints on allowable residue types and naturally supports extensions beyond the 20 canonical amino acids, including non-canonical variants, without requiring architectural modification.

To facilitate information flow between the continuous and discrete modalities, A-CODE's architecture adopts a shared transformer-based denoiser network built upon the noisy $X_t$ and $A_t$, with the final light-weight coordinate and type prediction heads for denoising:
\begin{equation}
    [\hat{X}_0, \text{Logits}] \gets \text{DenoiseNet}(X_t, A_t, \sigma_t, t, C),
\end{equation}
where $C$ is the conditional information in the binder design task. In this way, the model effectively learns the subtle yet crucial interdependence between the coordinate and type modalities, enabling simultaneous denoising during the iterative sampling. More details regarding the model architecture are provided in Appendix~\ref{suppl:arch}.

\subsubsection{Training Pipeline}
The training pipeline of \ourmodel utilizes a combination of continuous and discrete diffusion losses for coordinate-type co-design.
Let $X_0$ denote the extended clean coordinates where the coordinates of the dummy atoms are padded. The noisy coordinates are obtained via $X_t:=X_0 + \sigma_t\varepsilon$, where $\varepsilon\sim\mathcal{N}(0,1)$ is the standard Gaussian noise and $\sigma_t$ are defined according to the EDM scheduler~\cite{karras2022elucidating} used in the original AF3 model. Upon such noise structures, the original structure losses in the AF3 model are applied, which include the coordinate diffusion loss (EDM diffusion loss), a smooth LDDT loss, and additional auxiliary losses described in Appendix~\ref{suppl:train}. Following PXDesign, the structures of the conditional information are not fixed during training; instead, they are provided to the model as additional information within a unified learning framework. In this way, all structural losses also apply to the non-designable conditional structures. In practice, as with PXDesign, we observed that such information is learned quickly in the early stages of model training, with little error.

Let $A_0$ denote the extended clean atom types where the special \texttt{DMY} tokens are padded. The noisy atom types are obtained via the standard mask diffusion, where all the \texttt{XPA} tokens serve as the mask. Following~\citet{sahoo2024simple}, each designable atom type (i.e., not backbone and not condition in binder design) has a possibility of $\kappa_t$ of being turned into the corresponding \texttt{XPA} token, where we choose a square scheduler $\kappa_t=1-(1-t)^2$ following \citet{gat2024discrete}. The standard mask diffusion loss is then applied to designable atoms as 
\begin{equation}
    \mathcal{L}_\text{type}=\mathbb{E}_{t}\left[-\frac{\dot{\kappa}_t}{1-\kappa_t}\text{CE}(A_0,\phi_\theta(A_t,t))\right],\label{eqn:discrete_loss}
\end{equation}
where $\phi_\theta$ is the additional atom-level logit head that outputs atom type logits. We weigh the cross-entropy loss for the dummy atoms at different positions according to their natural frequencies to balance training. 
\subsubsection{Sampling Pipeline and Improvement}

As a unified co-design model, \ourmodel simultaneously generates discrete atom types and continuous atom coordinates in a single diffusion process, in contrast to two-stage approaches that decouple structure and sequence generation. For the coordinate modality, we adopt the same EDM solver as used in AF3. For the discrete modality, we follow the reparameterized discrete diffusion model (RDM)~\cite{zheng2023reparameterized}, which allows previously unmasked tokens to be remasked and has been shown to improve sampling robustness.
Specifically, based on the scheduler-determined mask proportion, the least-confident designable atom types are remasked. We further adopt a conservative decoding strategy in which a token is remasked if its current prediction is inconsistent with the previous step (see Appendix~\ref{suppl:sample} for details).

A critical challenge in fully atomic co-design is that prematurely committing side-chain conformations can introduce steric clashes, which in turn hinder the formation of a correct global fold. To mitigate this issue, we introduce a \textbf{side-chain denoising lag} strategy. Specifically, backbone atoms follow the standard noise schedule $\sigma_{\mathrm{bb}}(t)$, while side-chain atoms are assigned a lagged schedule $\sigma_{\mathrm{sc}}(t) = \eta \cdot \sigma_{\mathrm{bb}}(t)$ with $\eta > 1.0$ (we use $\eta = 1.5$), keeping side-chain atoms in a higher-noise regime during early denoising steps and biasing the model toward stabilizing the backbone topology before refining side-chain geometry. We find that applying this denoising lag at inference time alone is sufficient to improve unconditional monomer generation. For conditional binder design, we further apply the same lag during training, which improves optimization stability and downstream design performance. Unless otherwise stated, side-chain denoising lag is enabled during inference for all tasks, and additionally during training for binder design.

To ensure numerical stability when mixed-noise structures are passed to the denoiser, we apply a variance correction to the input coordinates:
\begin{equation}
\hat{X}_{\mathrm{input}} = X \cdot
\sqrt{\sigma_{\mathrm{bb}}^2 + \sigma_{\mathrm{data}}^2}\Big/
\sqrt{\sigma_{\mathrm{atom}}^2 + \sigma_{\mathrm{data}}^2},
\end{equation}
where $\sigma_{\mathrm{atom}}$ denotes the noise level assigned to a given atom (either $\sigma_{\mathrm{bb}}$ or $\sigma_{\mathrm{sc}}$). This normalization ensures that the effective input variance remains consistent with the training distribution, while deferring side-chain commitment until a reliable backbone scaffold has emerged.


\begin{algorithm}[ht]
\caption{Sampling From  \ourmodel}
\label{alg:sampling}
\begin{algorithmic}[1]
\STATE Calculate shareable conditional features $C$.
\STATE \textbf{Initialize:} Noisy coordinates $X_1 \sim \mathcal{N}(0, \sigma_{\text{max}}^2 \mathbf{I})$ and atom types $A_1 \gets \text{masked (\texttt{XPA})}$.
\STATE $\sigma_\text{bb}(t) \gets \text{EDM schedule from } \sigma_{\text{max}} \text{ to } 0$. $\sigma_\text{sc}(t) \gets \eta \cdot \sigma_\text{bb}(t)$.
\FOR{$t \gets 1, 1-1/N, \dots, 1/N$}
    \STATE $\sigma_\text{bb}, \sigma_\text{sc} \gets$ noise levels at current step $t$.
    \STATE $\lhd$ \textbf{Phase 1: Variance Normalization}
    \FOR{each atom $i$}
        \STATE $\sigma_i \gets \sigma_\text{bb}$ if $i \in \text{backbone}$, else $\sigma_\text{sc}$. $w_i \gets \sqrt{\sigma_\text{bb}^2 + \sigma_\text{data}^2} / \sqrt{\sigma_i^2 + \sigma_\text{data}^2}$.
    \ENDFOR
    \STATE $\lhd$ \textbf{Phase 2: Unified Multimodal Prediction}
    \STATE $\tilde{X}_t \gets X_t \odot \mathbf{w}$ \COMMENT{Element-wise scaling}
    \STATE $[\hat{X}_0, \text{Logits}] \gets \text{DenoiseNet}(\tilde{X}_t, A_t, \sigma_\text{bb}, t, C)$
    \STATE $\hat{X}_0 \gets \hat{X}_0 \oslash \mathbf{w}$ \COMMENT{Rescale predicted coordinates back}
    \STATE $\lhd$ \textbf{Phase 3: Coordinate \& Type Update}
    \STATE $X_{t-1/N} \gets \text{EDM\_Step}(X_t, \hat{X}_0,\sigma)$ using per-atom noise levels $\sigma_i$.
    \STATE $A_{t-1/N} \gets \text{RDM\_Step}(\text{Logits}, A_t,t)$.
\ENDFOR
\STATE 
Strip dummy atoms and infer residue types $S_0$ from $A_0$ via exact atom-name matching.
\STATE \textbf{return} $X_0, A_0, S_0$.
\end{algorithmic}
\end{algorithm}

Throughout sampling, no residue-level information is used or predicted. Residue identities are inferred only after all atom-level predictions have been fully denoised, by exact matching of the predicted atom types after removing dummy atoms. Although this procedure is sensitive to atom-type errors and assigns \texttt{UNK} when matching fails, we empirically observe a very low unknown residue rate of approximately $0.05\%$. The full sampling procedure is summarized in Algorithm~\ref{alg:sampling}, with additional details provided in Appendix~\ref{suppl:sample}. Notably, updates to the atom coordinates and atom types are performed in parallel using a single model forward pass at each diffusion step.

\subsection{Non-Canonical Amino Acid Modeling}\label{sec:ncaa}

All amino acids beyond the 20 standard proteinogenic ones are referred to as \emph{non-canonical amino acids} (ncAAs). ncAAs are widely present in natural proteins, most commonly arising from post-translational modifications. Their chemical diversity, together with relatively low per-type occurrence frequencies, makes ncAAs difficult to model using residue-level labels. In the PDB database, we identify 890 distinct ncAA types, each corresponding to a unique chemical component dictionary (CCD) code. Despite this sparsity at the type level, ncAAs appear frequently in aggregate, with approximately 46k total occurrences in the PDB, underscoring the importance of modeling ncAAs, particularly for peptide-based therapeutics.

Most existing protein design models effectively ignore ncAAs by restricting generation to backbone atoms, thereby discarding the all-atom information critical for side-chain interactions. In contrast, \ourmodel naturally supports ncAA modeling due to its atom-centric formulation, which imposes no hard constraints on residue identities during atom-type decoding. 
To leverage the strong generative prior learned from large-scale canonical protein data, we finetune a pretrained \ourmodel on structures containing ncAAs rather than training a separate model from scratch. This is achieved by constructing an atom-name mapping that assigns each non-canonical atom to a canonical atom name based on chemical semantics, including element type and relative position (see Appendix~\ref{suppl:data} for details). Apart from this mapping, all other featurization and training components remain identical, allowing ncAA modeling to be incorporated without architectural modification. 

\section{Experimental Result}

To demonstrate the effectiveness and flexibility of \ourmodel in all-atom generative modeling, we conduct extensive experiments spanning unconditional protein generation and conditional binder design. We also introduce a new ncAA benchmark for our new setup, which, to the best of our knowledge, has not been explored in prior work.

Following PXDesign, \ourmodel adopts the general AF3 architecture, except that the pairformer stacks are discarded for efficiency, resulting in a total of 169M trainable parameters. We also follow PXDesign's datasets, which include all PDB entries weighted by their pLDDT scores, as well as the clustered and filtered AFDB, a distilled dataset containing AlphaFold-generated structures. The model is trained progressively with two separate training stages. In the first stage, single-chain datasets predominate, whereas multi-chain binder design datasets are upweighted in the second stage. For fair comparison, we follow PXDesign to use 1000 Euler steps for diffusion sampling. Any unidentified residues will be converted to \texttt{UNK} during postprocessing, and later converted to glycine when fed into ESMFold for designability evaluation.

\subsection{Unconditional Protein Generation Benchmark}
\newcolumntype{Y}{>{\centering\arraybackslash}X}
\newcolumntype{W}{>{\hsize=1.3\hsize}Y} 
\newcolumntype{S}{>{\hsize=0.76\hsize}Y} 

\begin{table}[ht]
\centering
\caption{Unconditional protein generation benchmark results. Best results are marked in \textbf{bold}.}\label{tab:uncond}
\resizebox{\linewidth}{!}{
\begin{tabular}{@{} l cccc cccc @{}}
\toprule
\multirow{2}{*}{Model} & \multicolumn{2}{c}{Codesignability/\%$\uparrow$} & \multicolumn{2}{c}{Designability/\%$\uparrow$} & \multicolumn{2}{c}{Sec. Struct./\%} & \multicolumn{2}{c}{Diversity$\uparrow$} \\ 
\cmidrule(lr){2-3} \cmidrule(lr){4-5} \cmidrule(lr){6-7} \cmidrule(lr){8-9}
 & All-Atom & CA-Only & PMPNN@1 & PMPNN@8 & $\alpha$ & $\beta$ & Str & Seq \\ \midrule
P(all-atom) [\citenum{qu2024p}] & 36.7 & 37.9 & 44.4 & 57.9 & 56 & 17 & 134 & 148 \\
APM [\citenum{chen2025all}] & 19.0 & 32.2 & 42.8 & 61.8 & 73 & 8 & 32 & 64 \\
PLAID [\citenum{lu2024generating}] & 11.0 & 19.2 & 23.8 & 37.6 & 44 & 14 & 25 & 38 \\
ProteinGenerator [\citenum{lisanza2025multistate}] & 9.8 & 17.8 & 42.8 & 54.2 & 78 & 5 & 12 & 28 \\
MultiFlow [\citenum{campbell2024generative}] & - & 73.2 & 81.4 & 94.0 & 72 & 12 & \textbf{209} & 247 \\ 
Protpardelle [\citenum{chu2024all}] & 8.8 & 35.2 & 43.8 & 56.2 & 65 & 14 & 10 & 37 \\ 
La-Proteina [\citenum{geffner2025proteina}] & 68.4 & 72.2 & 82.6 & 93.8 & 72 & 5 & 206 & 216 \\
La-Proteina (w/ tri) & 75.0 & 78.2 & 84.6 & 94.6 & 73 & 6 & 129 & 252 \\ \midrule
\ourmodel & \textbf{81.2} & \textbf{87.0} & \textbf{91.6} & \textbf{97.0} & 50 & 14 & 70 & 252 \\
\quad w/o side-chain lag & 72.6 & 77.6 & 91.6 & 97.0 & 50 & 13 & 65 & \textbf{262} \\
\bottomrule
\end{tabular}
}
\end{table}

\begin{figure}[ht]
\centering
\includegraphics[width=\linewidth]{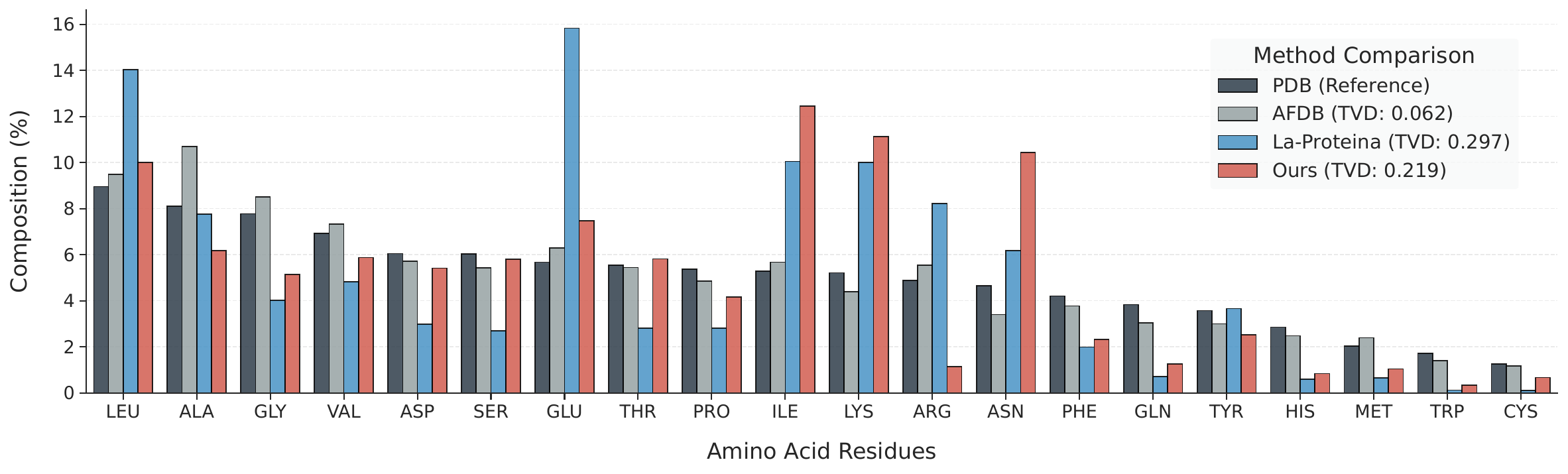}
\vspace{-1.5em}
\caption{Amino acid type distributions with total variation distance (TVD) to the PDB database.}
\label{fig:aa_dist}
\end{figure}

\begin{wraptable}{r}{0.52\textwidth}
\vspace{-1em}
\centering
\caption{Inference time (seconds) per sample across different lengths on a single NVIDIA A100-80G.}\label{tab:time}
\vspace{-0.5em}
\resizebox{\linewidth}{!}{
\begin{tabular}{@{}lccccc@{}}
\toprule
Model & 100 & 200 & 300 & 400 & 500 \\ \midrule
La-Proteina & 1.7 & 4.1 & 7.5 & 12.0 & 17.6 \\
La-Proteina w/ tri & 7.7 & 28.5 & 65.7 & 168.8 & 182.0\\
Multiflow & 23.5 & 25.8 & 30.5 & 44.9 & 61.0 \\
A-CODE & 2.9 & 5.2 & 7.8 & 10.9 & 14.6 \\ \bottomrule
\end{tabular}
}
\end{wraptable}
For unconditional protein design, we follow La-Proteina and draw 100 samples for each length in $\{100,200,300,400,500\}$. The metrics include:
\textbf{Codesignability}, defined as the percentage of the samples whose self-consistency RMSD (scRMSD) between the generated structure and the predicted structure from the ESMFold~\cite{lin2023evolutionary} on the generated sequence is less than 2.0 Å. The all-atom and C$\alpha$-only metrics indicate the atom sets used for the scRMSD calculation.
\textbf{Designability}, defined as the percentage of the samples whose scRMSD between the generated structure and the predicted structure from the PMPNN-redesigned sequence folded by ESMFold is less than 2.0 Å. For PMPNN@8, eight candidates are generated by the PMPNN, and the best one with the minimal scRMSD is selected, whereas only one sequence is generated for PMPNN@1.
\textbf{Diversity}, defined as the number of designable clusters in samples, using either the sequence or structure.
\textbf{Novelty}, defined as the highest TM-align scores against the target database (PDB or AFDB). A more detailed description of these metrics and the evaluation pipeline is available in Appendix~\ref{suppl:eval}.

The experimental results are summarized in Table~\ref{tab:uncond}. 
\ourmodel consistently generates high-quality designs, achieving a significant 6-9\% improvement in both designability and codesignability relative to the second-best La-Proteina, which uses the expensive triangular update. 
Importantly, side-chain lagging plays a crucial role in maintaining the consistency between the continuous and discrete modalities. Compared to consistently high designability, the lower codesignability indicates a mismatch between the generated structures and sequences. This suggests that deferring side-chain commitment is particularly important for joint structure–sequence generation. Our side-chain lagging effectively mitigates this issue by placing greater emphasis on the crucial backbone positions, thereby allowing the discrete types to be more accurately deciphered.

For inference time, \ourmodel demonstrates a 2.5-12x speedup over the second-best, La-Proteina (see Table~\ref{tab:time}), whereas La-Proteina, with a similar runtime, must discard the triangular update, resulting in even worse performance. Together, \ourmodel's superior designability and significant sampling efficiency support the effectiveness of our proposed fully-atomic unified framework.

We further provide an analysis of the distribution of the 20 canonical amino acids in the generations, as demonstrated in Figure~\ref{fig:aa_dist}. \ourmodel achieves a lower total variation distance (TVD) from naturally occurring proteins in the PDB database, indicating better fitness to the overall amino acid type distribution, despite a seemingly more difficult target: exact atom-name matching. Interestingly, \ourmodel's generated amino acid type distribution captures the low-frequency amino acid (\texttt{HIS,MET,TRP,CYS}) much better than La-Proteina, probably due to our position-wise weighting scheme to balance the residue distribution in an implicit way. La-Proteina, on the other hand, relies on latent codes to decode the residue type and thus has an inductive bias towards more frequent amino acids.
A more comprehensive analysis of the atomic-level distances, bond angles, and torsion angles is provided in Appendix~\ref{suppl:results}, where \ourmodel generates physically plausible all-atom structures.

\subsection{Conditional Binder Design Benchmark}


\begin{table*}[htb]
\centering
\caption{Binder design benchmark on the 10 targets. Best results for each model type are in \textbf{bold}.}\label{tab:binder}
\vspace{-0.5em}
\resizebox{\linewidth}{!}{
\begin{tabular}{@{}llcccccccccc@{}}
\toprule
Model Type & Model & BHRF1 & H1 & IL17A & IL7RA & IR & PDL1 & SC2RBD & TNFa & TrkA & VEGFA \\ \midrule
\multirow{5}{*}{Two-Stage} & BoltzGen [\citenum{stark2025boltzgen}] & 14.56 & 19.21 & 0.31 & 12.41 & 22.34 & 14.26 & 0.69 & 0.88 & \textbf{28.65} & 7.69 \\
 & ODesign [\citenum{zhang2025odesign}] & 42.23 & 7.29 & 0.10 & 7.61 & 16.82 & 10.26 & 3.13 & 0.00 & 9.99 & 3.71 \\
 & RFDiffusion-3 [\citenum{butcher2025novo}] & 33.38 & 0.89 & 0.82 & 6.47 & 18.17 & 16.81 & 4.34 & 0.00 & 14.44 & 2.95 \\
 & PXDesign [\citenum{team2025pxdesign}] & \textbf{43.90} & 12.08 & 0.82 & \textbf{29.80} & 25.04 & \textbf{45.33} & 11.20 & 3.43 & 23.55 & \textbf{16.72} \\
 & \ourmodel (PMPNN) & 25.00 & \textbf{65.59} & \textbf{1.79} & 4.93 & \textbf{30.09} & 39.96 & \textbf{28.05} & \textbf{6.16} & 6.87 & 1.37 \\ \midrule
\multirow{2}{*}{One-Stage} & Protpardelle-1c [\citenum{lu2025conditional}] & 3.73 & 0.27 & 0.00 & 0.09 & 0.19 & 7.04 & 0.99 & 0.00 & 3.52 & 0.17 \\
& \ourmodel (Co-Design) & \textbf{22.87} & \textbf{55.71} & \textbf{1.24} & \textbf{4.05} & \textbf{41.67} & \textbf{28.70} & \textbf{37.50} & \textbf{7.57} & \textbf{3.70} & \textbf{0.96} \\ \bottomrule
\end{tabular}
}
\end{table*}

We benchmark the conditional generation ability of \ourmodel on a test set comprising 10 protein targets with diverse structural properties, as proposed in \citet{zambaldi2024novo}. In binder design, a target protein with known sequence and structure is fed in as the conditions for the generative model. Hotspot information of the binding site is often available as additional input.
For a unified evaluation, we follow PXDesign to use the filter from AF2-IG~\cite{watson2023novo}. For each different target, we sample 328-728 binders with lengths ranging from 80 to 130, following the same protocol in PXDesign, and report the percentage of designable samples as \textbf{Designability}. We evaluate both the model-generated co-designed sequence and the PMPNN-redesigned single sequence as separate variants. The designability of \ourmodel, together with existing baselines, is provided in Table~\ref{tab:binder}, where two-stage and one-stage (co-design) models are highlighted separately.

Used as a two-stage model, \ourmodel achieves the best designability on 5 out of the 10 tasks, outperforming all existing baselines. The one-stage co-design results, on the other hand, also consistently and significantly outperform the Protpardelle baseline across all 10 tasks, with up to a tenfold increase in the success rates. Notably, the performance gap between the co-designed sequence and the 1-pass PMPNN-redesigned sequence has narrowed significantly, and the co-design variant even outperforms the best two-stage model (which is \ourmodel itself) on two of the hardest tasks of SC2RBD and TNFa.
Enabling the lag consistently also improves success rates across 6 of 10 targets, as demonstrated in Figure~\ref{fig:binder_ab} in Appendix~\ref{suppl:results}.

The superior performance of \ourmodel demonstrates the necessity and benefit of incorporating atomic-level information to build a better design model: the co-design framework not only significantly pushes the performance boundary within one-stage design models but also benefits two-stage design by enabling more fundamental atomic-level interactions between modalities.

\subsection{Extension to ncAA Modeling}

While the two training stages above are effective, with an extremely low frequency of unknown residues at 0.05\%, they do not unlock the full power of \ourmodel for ncAA modeling. To achieve this, we first filter the entire PDB database and retain entries with at least one ncAA. This gives a total number of 23k single chains and 29k binder pairs where the second chain contains at least one ncAA. With the additional atom name mapping described in Section~\ref{sec:ncaa}, we finetuned the model trained after the two main stages using only these two datasets to emphasize its ncAA modeling ability. 

\begin{wrapfigure}{r}{.54\linewidth}
\vspace{-1em}
    \centering
    \begin{subfigure}{\linewidth}
        \centering
        \includegraphics[width=\linewidth]{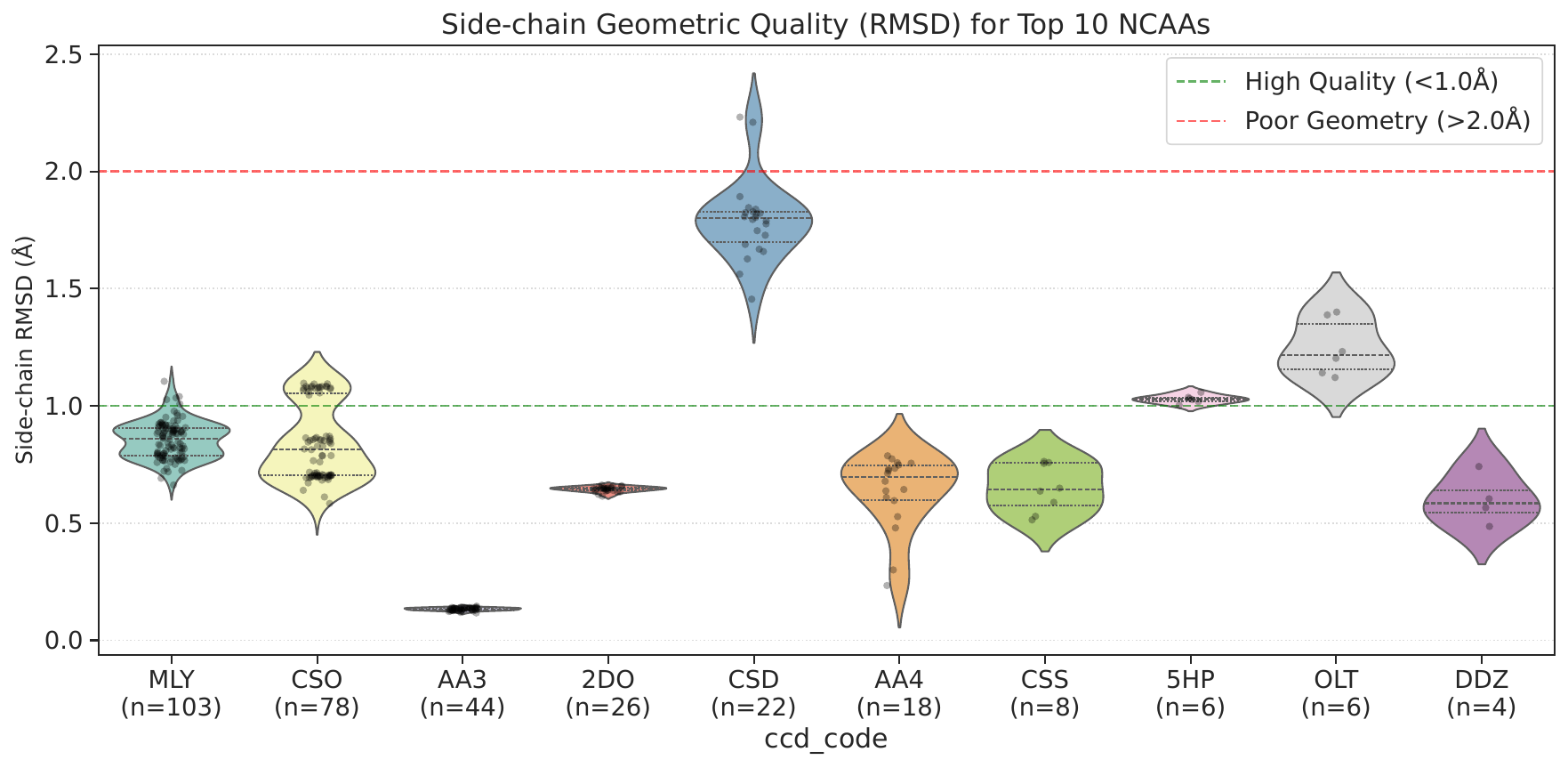}
        \caption{Side-chain RMSD.}
        \label{fig:ncaa_sc_rmsd}
    \end{subfigure}
    
    \begin{subfigure}{.56\linewidth}
        \centering
        \includegraphics[width=\linewidth]{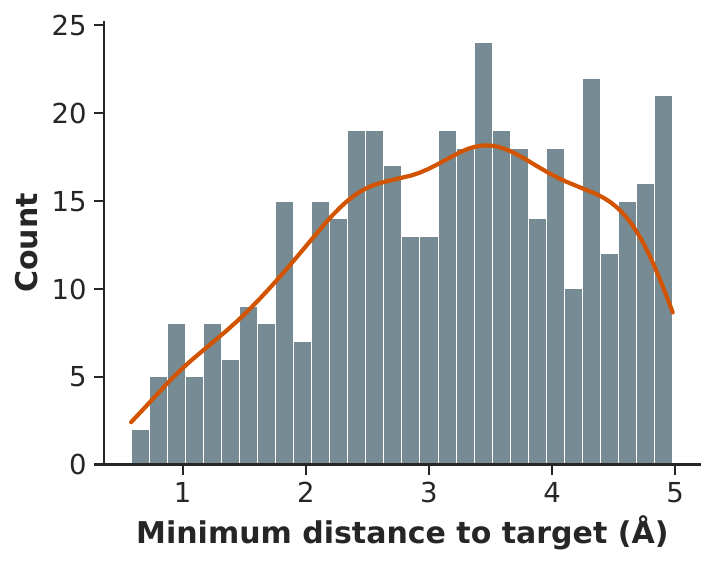}
        \caption{Minimum distance to target.}
        \label{fig:ncaa_min_dist}
    \end{subfigure}
    \hfill 
    \begin{subfigure}{.42\linewidth}
        \centering
        \includegraphics[width=\linewidth]{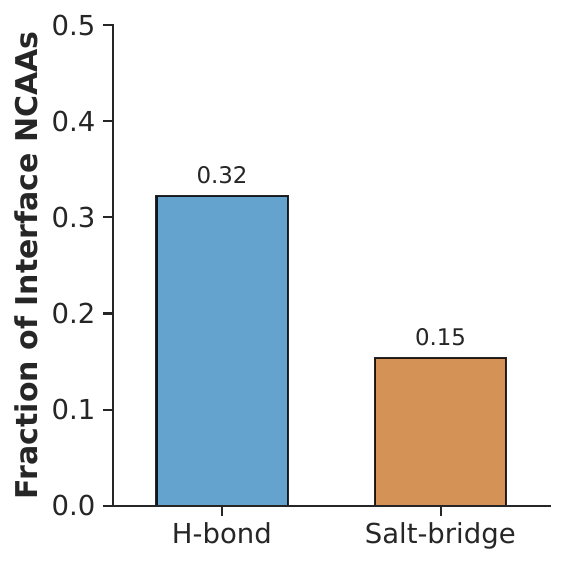}
        \caption{Interaction ratio.}
        \label{fig:ncaa_interface}
    \end{subfigure}
    \begin{subfigure}{\linewidth}
        \centering
        \includegraphics[width=\linewidth]{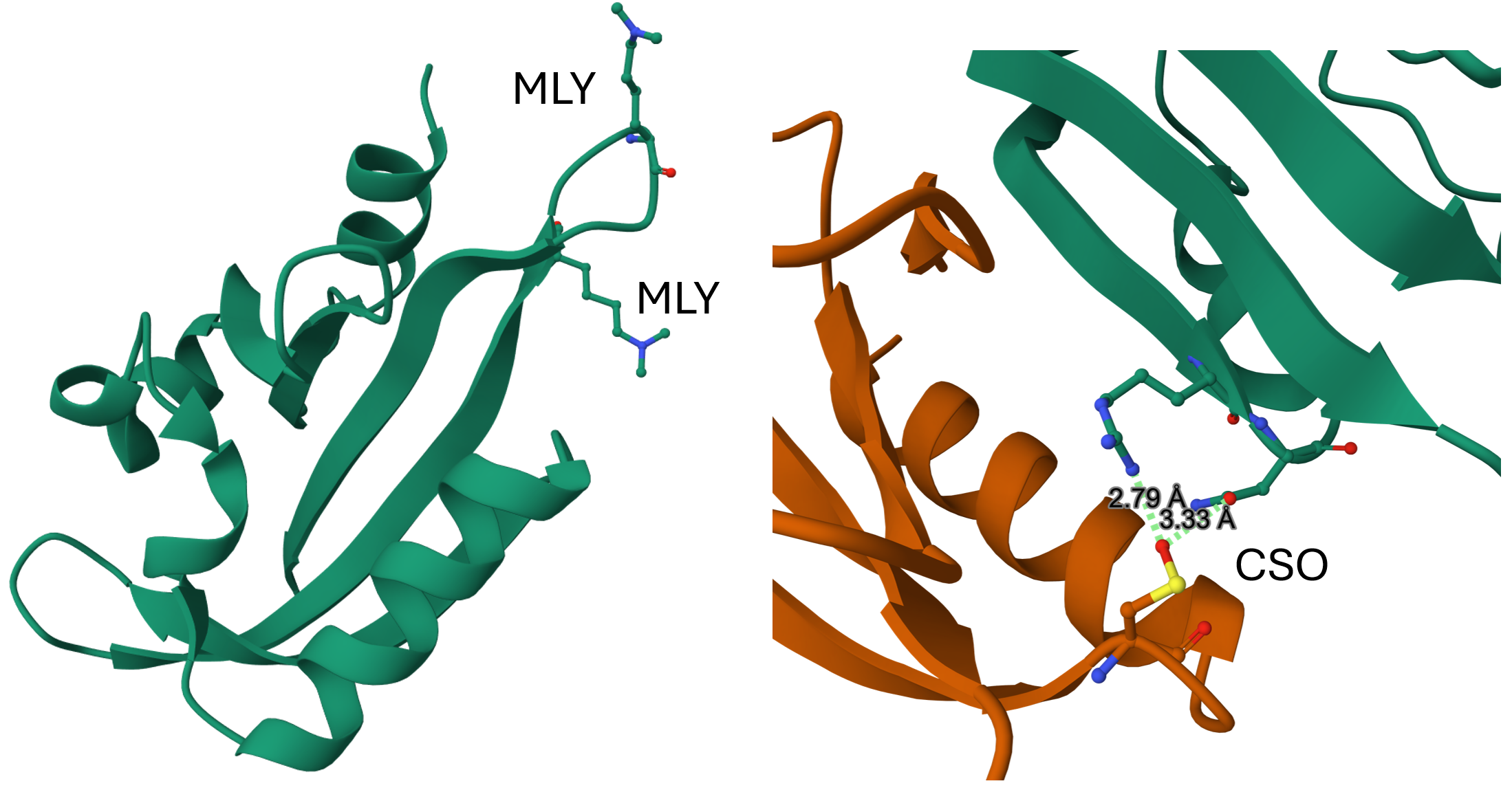}
        \caption{Generation examples.}
        \label{fig:ncaa_example}
    \end{subfigure}
    \caption{ncAA statistics analysis and generations.}
    \label{fig:ncaa}
\end{wrapfigure}
We provide a quantitative statistical analysis of the generated ncAAs in Figure~\ref{fig:ncaa}. Figure~\ref{fig:ncaa_sc_rmsd} illustrates the side-chain geometric quality for the top 10 most frequent ncAAs. The majority of generated structures are high-quality, with a side-chain RMSD under $1.0\text{ \AA}$, which demonstrates the robustness of \ourmodel in modeling complex chemical topologies. The distribution of the minimum distance from ncAAs to the target protein (Figure~\ref{fig:ncaa_min_dist}) shows a significant concentration between $2\text{ \AA}$ and $4\text{ \AA}$, the optimal range for biophysical interactions. Interface analysis in Figure~\ref{fig:ncaa_interface} reveals that 32\% of generated ncAAs participate in hydrogen bonding and 15\% form salt bridges. These statistics suggest that \ourmodel effectively captures the physicochemical preferences of ncAAs, leveraging their unique properties to stabilize binder-target complexes.

We further show representative generation results in Figure~\ref{fig:ncaa_example}. In the left panel, with the unconditional setup, two methyllysines (MLY) are successfully generated around the disordered loop region. This mimics the behavior of naturally occurring methyllysines in histones, in which methylation serves as a signal for gene regulation. Therefore, for signaling purposes, methyllysine residues are predominantly located in solvent-accessible regions or on the surface of protein complexes, as correctly captured by our \ourmodel generation.
On the right, we present the binder design setup, in which the target protein is specified as the condition. A hydroxycysteine (CSO) is generated at the interface instead of a cysteine, effectively expanding the sidechain length to promote sidechain interactions. Furthermore, with the modified geometry and the greater electronegativity of oxygen, two additional hydrogen bonds form with ARG and ASN residues in the target protein at appropriate distances, thereby reducing binding energy and stabilizing the complex.
In this way, we demonstrate the flexibility and feasibility of \ourmodel in ncAA modeling, offering a unified approach to all-atom protein generative modeling that can be extended to arbitrary biomolecules.


\section{Conclusion \& Limitation}
In this work, we propose \ourmodel, a fully atomic, one-stage protein co-design model with a unified multimodal diffusion framework for both discrete atom types and continuous atom coordinates. Our fully atomic design not only facilitates the synergistic generation of the two modalities but also enables natural extension to non-canonical amino acids. \ourmodel demonstrate superior generation quality, achieving new SOTA designability and codesignability on the unconditional benchmark and significantly better one-stage binder design performance across all ten tasks.

Despite its current superior performance over the baselines, we note limitations in \ourmodel. Most crucially, although the performance gap between the two-stage (with ProteinMPNN) and one-stage co-design models has been shrunk with our new state-of-the-art results, it remains non-negligible. Furthermore, our evaluations of the ncAA generations remain preliminary, as no widely acknowledged datasets or benchmarks have been established. In this way, we hope to inspire future design of protein generative models that naturally incorporate ncAAs and even ligands as a unified framework.

\section*{Broader Impacts}
We present \ourmodel for the fully atomic generation of protein structures and sequences within a unified multimodal diffusion framework. This advancement has broad implications for computational biology and drug discovery. By enabling the seamless incorporation of non-canonical amino acids (ncAAs) and eliminating the error propagation found in multi-stage pipelines, our model facilitates the design of more complex and functionally diverse protein therapeutics.

Our high-fidelity design tool lowers the barrier to entry for exploring novel protein modalities. While this fosters innovation in medicine and material science, it also necessitates a consideration of safety. The generalizability of our framework to arbitrary atomic compositions implies that it could potentially be applied to biological systems with unknown or hazardous properties. We advocate for the responsible release of model weights and are dedicated to ensuring the responsible use of our model for the positive benefit of society.

\bibliography{ref}
\bibliographystyle{plainnat}

\clearpage  
\newpage
\appendix
\centerline{\Large\bf Supplementary Material}

\section{Dataset Pipeline}\label{suppl:data}

In this section, we describe the details of our data pipeline for adapting PXDesign to an all-atom co-design model. Specifically, we discuss our procedure for the atom type (name) modality and our ncAA pipeline. 

\subsection{Canonical AA Pipeline}
\begin{table}[ht]
\centering
\caption{Padded atom names for the 20 canonical amino acids with frequency from \citet{beals1999aminoacidfrequency}.}\label{tab:aa}
\begin{tabular}{@{}ccc@{}}
\toprule
Amino Acid & Atom Names & Freq/\% \\ \midrule
ALA (A) & 
\verb|N  ,CA ,C  ,O  ,CB ,DMY,DMY,DMY,DMY,DMY,DMY,DMY,DMY,DMY| & 
7.4 \\
ARG (R) & 
\verb|N  ,CA ,C  ,O  ,CB ,CG ,CD ,NE ,CZ ,NH1,NH2,DMY,DMY,DMY| & 
4.2 \\
ASN (N) & 
\verb|N  ,CA ,C  ,O  ,CB ,CG ,OD1,ND2,DMY,DMY,DMY,DMY,DMY,DMY| & 
4.4 \\
ASP (D) & 
\verb|N  ,CA ,C  ,O  ,CB ,CG ,OD1,OD2,DMY,DMY,DMY,DMY,DMY,DMY| & 
5.9 \\
CYS (C) & 
\verb|N  ,CA ,C  ,O  ,CB ,SG ,DMY,DMY,DMY,DMY,DMY,DMY,DMY,DMY| & 
3.3 \\
GLN (Q) & 
\verb|N  ,CA ,C  ,O  ,CB ,CG ,CD ,OE1,NE2,DMY,DMY,DMY,DMY,DMY| & 
3.7 \\
GLU (E) & 
\verb|N  ,CA ,C  ,O  ,CB ,CG ,CD ,OE1,OE2,DMY,DMY,DMY,DMY,DMY| & 
5.8 \\
GLY (G) & 
\verb|N  ,CA ,C  ,O  ,DMY,DMY,DMY,DMY,DMY,DMY,DMY,DMY,DMY,DMY| & 
7.4 \\
HIS (H) & 
\verb|N  ,CA ,C  ,O  ,CB ,CG ,ND1,CD2,CE1,NE2,DMY,DMY,DMY,DMY| & 
2.9 \\
ILE (I) & 
\verb|N  ,CA ,C  ,O  ,CB ,CG1,CG2,CD1,DMY,DMY,DMY,DMY,DMY,DMY| & 
3.8 \\
LEU (L) & 
\verb|N  ,CA ,C  ,O  ,CB ,CG ,CD1,CD2,DMY,DMY,DMY,DMY,DMY,DMY| & 
7.6 \\
LYS (K) & 
\verb|N  ,CA ,C  ,O  ,CB ,CG ,CD ,CE ,NZ ,DMY,DMY,DMY,DMY,DMY| & 
7.2 \\
MET (M) & 
\verb|N  ,CA ,C  ,O  ,CB ,CG ,SD ,CE ,DMY,DMY,DMY,DMY,DMY,DMY| & 
1.8 \\
PHE (F) & 
\verb|N  ,CA ,C  ,O  ,CB ,CG ,CD1,CD2,CE1,CE2,CZ ,DMY,DMY,DMY| & 
4.0 \\
PRO (P) & 
\verb|N  ,CA ,C  ,O  ,CB ,CG ,CD ,DMY,DMY,DMY,DMY,DMY,DMY,DMY| & 
5.0 \\
SER (S) & 
\verb|N  ,CA ,C  ,O  ,CB ,OG ,DMY,DMY,DMY,DMY,DMY,DMY,DMY,DMY| & 
8.1 \\
THR (T) & 
\verb|N  ,CA ,C  ,O  ,CB ,OG1,CG2,DMY,DMY,DMY,DMY,DMY,DMY,DMY| & 
6.2 \\
TRP (W) & 
\verb|N  ,CA ,C  ,O  ,CB ,CG ,CD1,CD2,NE1,CE2,CE3,CZ2,CZ3,CH2| & 
1.2 \\
TYR (Y) & 
\verb|N  ,CA ,C  ,O  ,CB ,CG ,CD1,CD2,CE1,CE2,CZ ,OH ,DMY,DMY| & 
3.3 \\
VAL (V) & 
\verb|N  ,CA ,C  ,O  ,CB ,CG1,CG2,DMY,DMY,DMY,DMY,DMY,DMY,DMY| & 
6.8 \\ \bottomrule
\end{tabular}
\end{table}

In a standard PDB file, each of the 20 canonical amino acids has a unique and canonical ordered list of atom names. Such a list always starts with the 4 backbone atoms \texttt{N,CA,C,O} followed by the other side-chain atoms (if any). We note that identification based on element type and position to distinguish among amino acids may be ambiguous. For example, both LEU and ILE are identified by the sequence \texttt{N,C,C,O,C,C,C,C} and both PRO and VAL are identified by \texttt{N,C,C,O,C,C,C}. The atom names list is then padded with the special token \texttt{DMY} to make an atom14 representation for each amino acid as the target discrete modality. See Table~\ref{tab:aa} for a complete list of the padded names for all 20 canonical amino acids. For all amino acids, the input atom type is always fixed to be
\begin{equation}
    A_0:=\texttt{N,C,CA,O,XPA0,XPA1,XPA2,XPA3,XPA4,XPA5,XPA6,XPA7,XPA8,XPA9},
\end{equation}
where the 10 special \texttt{XPA} tokens serve as the mask token in discrete diffusion with additional positional embedding. The four backbone atoms are always fixed in both the forward noising process and the reverse sampling process to facilitate training. In this way, together with the \texttt{UNK} token for unknown atoms, we obtain a total number of 40 atom type classes, as shown in Table~\ref{tab:atom_names}. 

\begin{table}[ht]
\centering
\caption{Atom names used in \ourmodel. A total of 40 types are used.}\label{tab:atom_names}
\begin{tabular}{@{}lcc@{}}
\toprule
Atom Category & Atom Names & Count \\ \midrule
Backbone Atom & \verb|N,CA,C,O| & 4 \\ \midrule
Sidechain Atom & \begin{tabular}[c]{@{}c@{}}\verb|CB,CG,CG1,CG2,CD,CD1,CD2,CE,CE1,CE2,CE3,CZ,CZ2,CZ3,CH2,|\\ \verb|ND1,ND2,NE,NE1,NE2,NZ,NH1,NH2,|\\
\verb|OG,OG1,OD1,OD2,OE1,OE2,OH,OXT,SG,SD|\end{tabular} & 33 \\ \midrule
Special Atom & \begin{tabular}[c]{@{}c@{}}\verb|DMY| (dummy atom in the target)\\ \verb|XPA| (design atom in the input)\\ \verb|UNK| (unknown atom)\end{tabular} & 3 \\ \bottomrule
\end{tabular}
\end{table}

During training, the dummy atoms are randomly sampled from a Gaussian distribution with a unit variance centered at the CA atom. In this way, we can provide an additional supervising signal for the model to learn the approximate position of the amino acid while offering robustness to potential noise compared to a deterministic target. During sampling, the dummy atoms are not deleted until the final postprocessing (matching) stage, ensuring a consistent model input. After the diffusion sampling stage finalizes, the postprocessing stage identifies the residue type based on the final atom types predictions in the procedure described in Algorithm~\ref{alg:match}. It can be seen that the exact-matching scheme requires both the generation's positions and atom types to be identical for a successful match.

\begin{algorithm}[ht]
\caption{Residue Type Matching}\label{alg:match}
\begin{algorithmic}[1]
\STATE \textbf{Input:} Predicted atom types $A_0$.
\FOR{Each residue}
    \STATE Strip any dummy atom \texttt{DMY}.
    \STATE Concatenate the remaining atom names as the unique identifier.
    \STATE Matching the identifier for the 20 canonical amino acids.
    \STATE If some amino acid matches, the residue type is assigned; else, assign \texttt{UNK}.
    \STATE If more than one atom shares the same atom name (which only happens in \texttt{UNK} or ncAAs after atom name mapping), modify the repetitive names by appending or increasing the last numerical count in the name. This ensures that the output PDB file is legal.
\ENDFOR
\STATE \textbf{return:} $S_0$. 
\end{algorithmic}
\end{algorithm}

The featurization generally follows the original AF3 framework, with some modifications on the designable residues/atoms to prevent data leakage. Specifically, \texttt{restype} for designable residues is assigned a special \texttt{XPB} token. Any MSA or template information is disabled, with features set to zero as inputs. Additionally, the atom-level CCD reference information for the designable residues is replaced with that for glycine, which provides only backbone information. For conditional generation in binder design, additional ground-truth coordinates, residue types, and reference information are still available in a separate conditional track, enabling the model to recover the target protein structure. For binder pairs, a hotspot mask will be randomly sampled as additional information to enable hotspot-conditioned binder design during inference.

In the first two stages of the training, we follow standard practice from previous work by first ignoring any ncAA and substituting them with glycine. That is, we only extract the positions and atom names of the four backbone atoms \texttt{N,CA,C,O} and discard any sidechain information as if it were a glycine. In this way, we aim to minimize the occurrence of \texttt{UNK} at these stages, leaving the ncAA modeling to the third finetuning stage. 

The datasets used in the first stage follow the original setup in PXDesign, where we have a combination of the experimentally determined PDB, split into single-chain and binder-pair splits, as well as the distillation dataset of AlphaFold-predicted structures in AlphaFold Database (AFDB), which contains only single-chain structures. The single-chain PDB split contains 433,318 data points, clustered into 36,955 unique clusters. The PDB complex split contains 266,369 data pairs, clustered into 26,394 unique clusters. The AFDB split, after filtering and clustering, contains 8,403,792 single-chain data points.

\subsection{ncAA Pipeline}

For non-canonical amino acids, to enable finetuning without expanding the current atom type vocabulary, we first propose a semantic atom name mapping that assigns a \emph{chemically} closest atom name to the atom names in ncAAs. As the atom names in the PDB file often follow a naming convention, where the first character indicates the element type, the second character indicates the relative position (A, B, G, D, etc.), and the last character indicates the numbering, we may provide more semantically important chemical information in addition to pure element types. The detailed atom name mapping scheme is described in Algorithm~\ref{alg:name_map}. We note that the unknown atoms only occur when an element other than C, N, O, or S is present, such as P in phosphorylation.

\begin{algorithm}[ht]
\caption{ncAA Atom Name Mapping}\label{alg:name_map}
\begin{algorithmic}[1]
\STATE \textbf{Input:} Atom name.
\STATE If the name matches a standard one, return the name.
\STATE Else, if the element type and positions match, return the first available name with the same first two characters.
\STATE Else, if the element type matches, return the last available position of that element.
\STATE Else, return the unknown atom type.
\STATE \textbf{return:} Mapping canonical atom name. 
\end{algorithmic}
\end{algorithm} 

We also ensure the four backbone atoms are always the first four atoms in the ncAA, where the other sidechain atoms follow the original ordering in the PDB. For compatibility with the atom14 representation, any ncAA with more than 14 heavy atoms will be truncated to keep only the first 14. Other parts of the ncAA pipeline are the same as the standard pipeline, demonstrating the flexibility and transferability of our \ourmodel framework.

In principle, it is always possible to incorporate the ncAA atom names in addition to the 40 standard AA atom name vocabulary, but this will lead to unnecessary expansion of the vocabulary with extremely low-frequency classes. Nonetheless, we do note two limitations that can be potentially improved by training a new model: a) Although the unknown atom name is uncommon after mapping, they do exist for atom types other than C, N, O, or S. Adding some additional element types to the vocabulary will mitigate this issue. b) Although the atom14 representation suffices for the 20 standard AA and most ncAAs, there are ncAAs with more than 14 atoms. We truncate those ncAAs with more than 14 heavy atoms. Using larger atom representations will mitigate this issue at the cost of higher computational cost. We will leave more comprehensive explorations for future work.

The ncAA dataset is constructed by filtering the whole PDB entries and keeping only the entries with at least one ncAA, with the exceptions that a) the already existing \texttt{UNK} residues are regarded as alanine and not ncAA, and b) any residues without the four backbone atoms are not regarded as ncAA (e.g., the amino group \texttt{NH2}, the acetyl group \texttt{ACE}, or the methylamino group \texttt{NME}). Note that the D-amino acids (or other enantiomers/diastereomers) are still kept as ncAAs. Other non-protein components, such as DNA, RNA, or ligand, are filtered away. Following the PXDesign dataset splitting scheme, we also construct separate splits for the single-chain and binder ncAA datasets, where for the latter, the second chain is fixed as the binder (designable) and should contain at least one ncAA. After filtering, we obtain 22632 single-chain data and 29252 binder pairs. The whole dataset contains 890 distinct ncAA types, out of which 263 have more than 14 heavy atoms, and the largest ncAA contains 61 heavy atoms, where a coenzyme A is attached.

\section{Training Pipeline}
In this section, we provide additional details for training \ourmodel in the three stages.

\subsection{Model Specification}\label{suppl:arch}
\begin{figure}[ht]
\centering
\includegraphics[width=\linewidth]{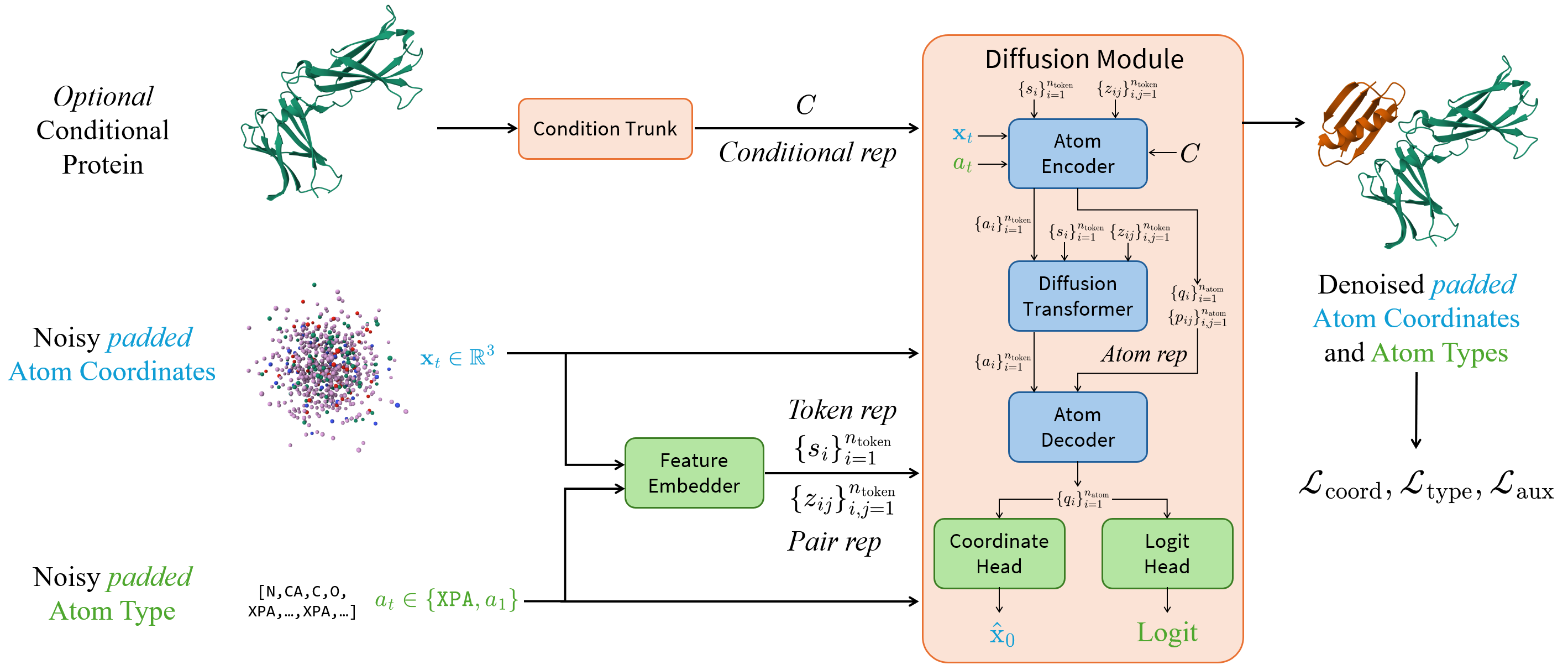}
\vspace{-1em}
\caption{Model architecture of A-CODE. For unconditional generation, the condition $C$ will be generated with placeholders and mask features. Note how the atom types are added as additional input features and output predictions.}
\label{fig:arch}
\end{figure}

Following PXDesign, we do not use the full AF3 architecture and discard all pairformer layers to improve efficiency, resulting in a small model with 169M trainable parameters for efficient generative modeling. The overview of the A-CODE model architecture is provided in Figure~\ref{fig:arch}.
The conditional trunk and the feature embedder generally follow the original AF3 design, in which additional node features, such as position embeddings and chain IDs, are fed into the module along with noisy coordinates and types. The conditional trunk offers additional structure information as soft constraints. For unconditional generation, these features will be generated from placeholders of \texttt{XPB} residue types (noted as X in Figure~\ref{fig:model}) to prevent information leakage. The embedder generations token representations $\{s_i\}_{i=1}^{n_\text{token}}$ and token-wise pair representations $\{z_{ij}\}_{i,j=1}^{n_\text{token}}$ via reading out over atom features. These features, without requiring expensive pairformer stacks for refinement, are directly fed into the diffusion module.

The diffusion module consists of an atom encoder, a diffusion transformer, and an atom decoder, followed by the final prediction heads for denoising. The atom encoder takes the token-level single and pair representations, the conditional features, and the atomic-level noisy coordinates and types as inputs. The token-level representations are directly fed into the diffusion transformer for refinement, whereas the atom-level single and pair representations $\{q_i\}_{i=1}^{n_\text{atom}},\{p_{ij}\}_{i,j=1}^{n_\text{atom}}$ are fed in a skip-connection fashion directly to the atom decoder. In this way, the overall computational cost will not increase dramatically when adopting the atom14 representation, which has more atoms but the same number of tokens. The output of the atom decoder is the atom-level representation, upon which the denoised coordinates $\hat{\mathbf{x}}_0$ and the atom type logits are predicted following separate heads.

We generally follow the hyperparameter selection used in PXDesign, except that the query sizes in the atom encoder and decoder are increased to 128 to accommodate more atoms. We refer interested readers to the original AF3 for the details on each module and PXDesign for detailed hyperparameters.

\subsection{Training Specification}\label{suppl:train}
During training, the noisy interpolation data are generated using a combination of continuous and discrete diffusion. Specifically, for the continuous atom coordinates, we follow AF3 to use the EDM scheduler~\cite{karras2022elucidating} as:
\begin{equation}
    \mathbf{x}_t=\mathrm{random\_aug}(\mathbf{x}_0)+\sigma_t,
\end{equation}
where $\mathrm{random\_aug}$ denotes the random rotation augmentation and $\sigma_t\sim \sigma_\text{data}\cdot\exp(-1.2+1.5\mathcal{N}(0,1))$ with $\sigma_\text{data}=16$. For the structure losses, we generally follow the original ones in AF3. The major diffusion loss is a combination of the weighted MSE loss and the smooth local distance difference test (LDDT) loss with a cutoff distance of 15 Å:
\begin{equation}
    \mathcal{L}_\text{coord}=\frac{\sigma_t^2+\sigma_\text{data}^2}{(\sigma_t+\sigma_\text{data})^2}\mathcal{L}_\text{MSE}+\mathcal{L}_\text{slddt},
\end{equation}
where the MSE is calculated between the denoiser prediction $\hat{X}_0$ and the ground truth aligned to the prediction $X_0^\text{align}$. The smooth LDDT loss is defined as a smooth distance difference between the ground truth and the prediction structures as:
\begin{equation}
    \mathcal{L}_\text{slddt}=1-\frac{1}{|D|}\sum_{(i,j)\in D}F(|\hat d_{ij}-d_{ij}|),
\end{equation}
where $d_{ij}=\|\mathbf{x}^\text{GT}_i-\mathbf{x}^\text{GT}_j\|$ denotes the distance between the ground truth $X_0$ and $\hat d_{ij}=\|\hat{\mathbf{x}}_i-\hat{\mathbf{x}}_j\|$ for the prediction $\hat X_0$. The summation is over the set $D=\{(i,j)|i\ne j,d_{ij}<15\text{Å},1\le i,j\le n_\text{atom}\}$ of atom pairs $(i,j)$, and the soft kernel is defined as 
\begin{equation}
    F(\delta)=\frac{1}{4}\left[\sigmoid\left(\frac{1}{2}-\delta\right)+\sigmoid\left(1-\delta\right)+\sigmoid\left(2-\delta\right)+\sigmoid\left(4-\delta\right)\right].
\end{equation}
Additionally, auxiliary losses for predicting denoised structure distogram bins, LDDT (pLDDT), aligned error (pAE), and distance error (pDE) are also leveraged for all atoms (including conditions) following the original setup in AF3. We refer interested readers to the original AF3 paper~\cite{abramson2024accurate} for more details.

For the discrete atom type, we adopt the standard mask diffusion model (MDM), where the noisy discrete atom types are obtained as:
\begin{equation}
a_t=
\begin{cases}
    a_0, &\text{w.p. }\kappa(t),\\
    \texttt{XPA}, &\text{w.p. }1-\kappa(t),
\end{cases}
\end{equation}
where we follow \citet{gat2024discrete} to use the square scheduler $\kappa(t)=1-(1-t)^2$ (note that we use the diffusion convention such that $t=0$ is the clean data). For better consistency between the two modalities, we use a coupled scheduler where $t=F^{-1}(\sigma_t)$ is inferred from the inverse CDF $F^{-1}$ of $\sigma_t$.

The discrete diffusion loss is demonstrated in Eq.~\ref{eqn:discrete_loss} and is calculated on designable atoms (designable residue and not backbone). The discrete diffusion loss is weighted by the natural occurrence frequencies of the 20 canonical amino acids to prevent data imbalance between the dummy and non-dummy targets. We follow the statistics in Table~\ref{tab:aa} to adjust the positive weight (non-dummy) as if there is an equal number of dummy and non-dummy atoms at each position. As an example, the \texttt{XPA0} position is 92.6\% non-dummy, whereas the \texttt{XPA8} and \texttt{XPA9} positions only have 1.3\% non-dummy, so the dummy cross-entropy losses on these positions are weighted by $12.51$ and $0.013$, respectively.

\begin{algorithm}[ht]
\caption{Training Step for A-CODE}\label{alg:train}
\begin{algorithmic}[1]
\STATE Calculate conditional features $C$.
\STATE Sample $\sigma_t\sim \sigma_\text{data}\cdot\exp(-1.2+1.5\mathcal{N}(0,1)),t=F^{-1}(\sigma_t)$.
\STATE $X_t\gets \text{Noise\_EDM}(X_1,\sigma_t)$.
\STATE $A_t\gets \text{Noise\_MDM}(X_1,\kappa(t))$.
\STATE $[\hat{X}_0, \text{Logits}] \gets \text{DenoiseNet}(X_t, A_t, \sigma_t, t, C)$
\STATE Calculate the continuous diffusion loss $\mathcal{L}_\text{coord}(\hat{X}_0,X_0,\sigma_t)$, the discrete mask diffusion loss $\mathcal{L}_\text{type}(\text{Logits}, A_0,\kappa(t))$ and other auxiliary losses.
\STATE Optimize the model parameters.
\end{algorithmic}
\end{algorithm}

The datasets we used for training \ourmodel in the first two stages are the same, except that the binder split was upweighted in the second stage. In the first stage, the dataset weights are 1.4, 0.004, and 3.25 for the PDB single-chain, PDB complex, and AFDB, respectively; in the second stage, they are 0.4, 0.2, and 0.8.
In the first two stages, ncAAs were discarded, and the ncAA datasets were not used. In the third stage, we used only the two ncAA datasets for finetuning to highlight the ncAA modeling ability of \ourmodel, with weights of 1.0 and 1.0. The model was trained with a batch size of 64, with 200k iterations in the first stage and 400k in the second. The model was then finetuned for another 200k iterations in the third stage.
At all stages, we used a cropping size of 640 residues, employing a mixture of spatial, sequential, and interface cropping approaches as described in the AF3 paper. We used a local batch size of 1 and a diffusion batch size of 8. A constant learning rate of $5\times 10^{-4}$ is applied for all training stages. All training was carried out on NVIDIA H20 GPUs.

\subsection{Sampling Specification}\label{suppl:sample}
The sampling component of the structure generally follows the original AF3 sampling algorithm, adapted from the EDM's second-order Heun solver. The inference-time noise schedule is defined as
\begin{equation}
    \sigma_t=\sigma_\text{data}\cdot\left(s_\text{max}^{1/p}+t(s_\text{min}^{1/p}-s_\text{max}^{1/p})\right)^p,
\end{equation}
where $s_\text{max}=160,s_\text{max}=4\times10^{-4},p=7$ and $t$ is uniformly distributed between $[0,1]$ with 1000 sampling steps. The detailed sampling algorithm is summarized in Alg.~\ref{alg:edm}, where each step advances the current noisy coordinates with a step size of $1/1000$ and a step scale $\eta=2.5$. As noted in Alg.~\ref{alg:sampling}, the noise schedule $\sigma_t$ here can be either $\sigma_\text{bb}$ or $\sigma_\text{sc}$ depending on the atom types.

\begin{algorithm}[ht]
\caption{Diffusion Sampling for Coordinates (EDM\_Step)}\label{alg:edm}
\begin{algorithmic}[1]
\REQUIRE Current time $t$, step size $\Delta t$, current coordinates $\mathbf{x}_t$, denoised coordinates $\hat{\mathbf{x}}$, schedules $\sigma_t,\sigma_{s}$ with $s=t-\Delta t$, noise scale $\lambda=1.003$.
\STATE $\gamma\gets 0.8\text{ if }\sigma_t >1\text{ else }0,\hat t\gets (\gamma+1)\sigma_s$
\STATE $\tilde{\mathbf{x}}_t\gets \mathbf{x}_t+\lambda\sqrt{\hat t^2-\sigma_{s}^2}\cdot \varepsilon,\varepsilon\sim\mathcal{N}(0,I)$
\STATE $\mathbf{x}_s\gets \tilde{\mathbf{x}}_t+\eta(\sigma_t-\hat{t})(\mathbf{x}_t-\hat{\mathbf{x}})/\hat t$
\end{algorithmic}
\end{algorithm}

For the discrete diffusion sampling of atom types, instead of using the standard mask diffusion sampling where modification of an already unmasked token is not allowed, we follow RDM~\cite{zheng2023reparameterized} to allow for a more flexible control of the sampling dynamics, which is described in Algorithm~\ref{alg:rdm}. 
As in training, we use the same square schedule for the discrete modality. Additionally, we employ the \emph{conservative} decoding scheme in RDM, where a token must be sufficiently confident to be unmasked, and less-confident predictions are remasked to allow modification. This is carried out iteratively by recording the previous logits as the confidence score and iteratively unmasking and remasking. An exception would be the last step, where we will force all the remaining masks to be unmasked. In practice, about 15\%-20\% of the tokens are unmasked during the last step, with the others already determined during the iterative process. In this way, our sampling scheme is fundamentally different from existing all-atom two-stage models, in which atom types are inferred solely from the final denoised structure after a single forward pass.

\begin{algorithm}[ht]
\caption{Discrete Diffusion Sampling for Types (RDM\_Step)}\label{alg:rdm}
\begin{algorithmic}[1]
\REQUIRE Current time $t$, step size $\Delta t$, previous selected logits $s_\text{prev}$ and token $a_\text{prev}$ at $t$, current selected logits $s_\text{cur}$ and token $a_\text{cur}$ at $t-\Delta t$, schedule $\kappa(\cdot)$.
\STATE If $t\le \Delta t$, unmask all the remaining masked positions with current tokens and return.
\STATE Sort the previous selected logits $s_\text{prev}$, mask $1-\kappa(t-\Delta t)$ proportion of the previous tokens with the lowest logits (low-confidence prediction).
\STATE For the current unmasked positions in the previous token, remask all the positions where $a_\text{prev}\ne a_\text{cur}$ (inconsistent prediction) or $s_\text{prev}> s_\text{cur}$ (less confident prediction).
\STATE For the current masked positions in the previous token, unmask all tokens with the current tokens where $a_\text{prev}= a_\text{cur}$ and $s_\text{prev}\le s_\text{cur}$.
\STATE Set $s_\text{prev}\gets s_\text{cur},a_\text{prev}\gets a_\text{cur}$ for the next iteration.
\end{algorithmic}
\end{algorithm}

\subsection{Additional Relative Work}\label{suppl:related}
We also include several notable unimodal protein design models as references. 
In principle, any structure-only design model can be cascaded with inverse folding models such as ProteinMPNN~\cite{dauparas2022robust} to form a co-design model, and any sequence-only design model can be cascaded with folding models like AF2/AF3 as a co-design model. 
For example, FrameFlow~\cite{yim2023fast} and FoldFlow~\cite{bose2023se} leverage Riemannian flow matching to learn the position and orientation of each amino acid's backbone atoms. Proteina~\cite{geffner2025proteina}, the predecessor of La-Proteina, is a CA-only structure design model. Other protein language models, such as InstructPLM~\cite{qiu2024instructplm} and DPLM~\cite{wang2024diffusion,wang2024dplm}, can also be adapted for protein co-design.
Within sequence design models, methods like InstructPLM~\cite{qiu2024instructplm} extend the power of large language models to protein design tasks by framing them as natural language tasks. On the other hand, models like DPLM~\cite{wang2024diffusion,wang2024dplm} rely purely on the amino acid sequences and structure tokens.
These models are generally posed to address different unimodal design tasks. Therefore, we do not include them in our experiments.

\section{Evaluation Pipeline}\label{suppl:eval}
In this section, we discuss the evaluation pipeline and metrics.

\subsection{Uncondtional Protein Generation}
The unconditional protein generation task benchmarks the general structure-sequence co-design ability of a generative model, highlighting the need for more complex practical tasks like binder design. We generally follow the setup in PXDesign to generate 128 sequences for each length in $\{100,200,300,400,500\}$ and then average across all generated samples to compute the metric. For co-design models in Table~\ref{tab:uncond}, we generally either copy the originally reported metrics or follow the recommended inference setup with the official checkpoints.

For comprehensive evaluation, we categorize the metrics we used into \emph{likelihood} and \emph{diversity} metrics. The likelihood metrics measure the quality of the generation, determining whether they resemble a \emph{natural} protein structure, sequence, or both. Such likelihood metrics include:

\paragraph{Designability} Designability describes whether a structure is \emph{foldable}, i.e., if there exists some amino acid sequence that folds into such a structure. In practice, ProteinMPNN~\cite{dauparas2022robust} (PMPNN) is used as a common inverse-folding model to design amino acid sequences, which are then fed into ESMFold~\cite{lin2023evolutionary} for folding prediction. In this way, designability is defined as the average proportion of the generated structures whose smallest scRMSD to the folded structures of the ProteinMPNN-design sequence is less than 2.0 Å. \textbf{PMPNN@1} and \textbf{PMPNN@8} denote the different numbers of candidate amino acid sequences generated by ProteinMPNN. Designability does not rely on the sequence modality generated by the model.

\paragraph{Codesignability} Codesignability describes whether a sequence-structure pair is feasible, offering another aspect of likelihood that also depends on the sequence modality and its coordination with the structure modality. Utilizing ESMFold on the generated sequence, codesignability is defined as the proportion of the generated structure-sequence pair whose scRMSD between the generated structure and the folded structure of the generated sequence is less than 2.0 Å. \textbf{All-Atom} and \textbf{CA-Only} denote the different alignment sets on either the all-atom structure or the CA atom only. The latter is generally easier. Due to the limitation of ESMFold, any generated unknown residue in this stage will be substituted with a glycine.

The diversity metrics measure the sample diversity of a generative model, preventing the model from overfitting only on high-likelihood data or memorizing the dataset, which includes:

\paragraph{Diversity} Diversity counts the number of distinct clusters in the generated samples, and a higher diversity score indicates a broader span of the generated samples. The clustering is obtained by running the \texttt{foldseek}~\cite{van2024fast} program, and the \textbf{Str}, \textbf{Seq}, and \textbf{Str+Seq} subtypes are configurable clustering types that consider either the structure or sequence modality during clustering.


\subsection{Binder Design}
The setup for the binder design follows the PXDesign paper~\cite{team2025pxdesign}, which uses the 10 protein targets proposed in \citet{zambaldi2024novo}, covering a wide range of structure diversity and exhibiting real-world biological significance.
Designability is the major metric used here, where we follow PXDesign to rely on the \textit{in silico} filter from AF2-IG~\cite{watson2023novo}, where a designable binder is defined as: a) interface predicted absolute error (ipAE) less than 10.85 Å, and b) interface predicted TM score (ipTM) greater than 0.5, and c) predicted local distance difference test (pLDDT) greater than 80\%, and d) binder bound/unbound RMSD less than 3.5 Å. AF2~\cite{jumper2021highly} is used to evaluate these metrics, in which residues in the binder chain are assigned a large offset to mimic multichain co-folding. Across different binder lengths for the same target, the success counts are summed to obtain the final result. 

\subsection{ncAA Analysis}\label{suppl:ncaa_eval}
The ncAA generation protocol follows the unconditional setup with 500 generations in total.
ncAAs are identified from \texttt{UNK} residues in the binder chain and matched to PDB Chemical Component Dictionary (CCD) entries using heavy-atom topology matching via graph isomorphism. The graph topology is inferred from the generated structures using RDKit\footnote[1]{\url{https://www.rdkit.org/}}, and only those whose graph topology and element types exactly match a CCD will be successfully identified.
Side-chain geometric quality was assessed by computing a side-chain RMSD to the corresponding CCD reference, excluding backbone atoms (N, CA, C, O) and taking the minimum RMSD over all symmetry-consistent atom mappings. Interface ncAAs were defined based on the minimum heavy-atom distance to the target protein (5 Å as a threshold), and interaction patterns were characterized using geometry-based proximity proxies, including hydrogen-bond-like (N/O/S atoms within 3.5~\AA) and salt-bridge-like (N--O atom pairs within 4.0~\AA) contacts. More quantitative results are available in Appendix~\ref{suppl:ncaa_result}.

\section{Additional Results}\label{suppl:results}

In this section, we provide additional experimental results to support a more comprehensive analysis of \ourmodel's generative performance.

\subsection{Additional Results for Unconditional Protein Generation}
\begin{table}[ht]
\centering
\caption{Atomic structural evaluation metrics of MAD and clashscore for \ourmodel and La-Proteina.}\label{tab:atomic}
\resizebox{\linewidth}{!}{
\begin{tabular}{@{}lcccccccccc@{}}
\toprule
\textbf{Metric (MAD)} & \textbf{N-CA} & \textbf{CA-C} & \textbf{C-N} & \textbf{C-O} & \textbf{N-CA-C} & \textbf{CA-C-N} & \textbf{C-N-CA} & \textbf{sc\_bond\_length} & \textbf{sc\_bond\_angle} & \textbf{clashscore} \\ \midrule
Ours & 0.0096 & 0.009 & 0.0185 & 0.0071 & 1.1618 & 1.521 & 1.6651 & 0.0188 & 1.8623 & \textbf{53.67} \\
LA-Proteina (notri) & 0.0062 & 0.0056 & 0.0184 & 0.0079 & 1.1781 & 1.3632 & 1.3565 & 0.0234 & 1.7959 & 70.06 \\
LA-Proteina (tri) & 0.0061 & 0.0061 & 0.0187 & 0.008 & 1.1577 & 1.9424 & 1.8639 & 0.0233 & 1.7862 & 57.50 \\ \bottomrule
\end{tabular}
}
\end{table}

To comprehensively evaluate the physical realism of the generated all-atom structures and explicitly validate the effectiveness of our fully atomic co-design approach, we conducted a rigorous quantitative analysis of atom-level physical metrics. The evaluations are aggregated across all generation lengths in the unconditional setup, conducted on structures generated by our method and compared against the baseline (La-Proteina) under identical settings. We focused on three primary aspects:
\begin{itemize}
    \item \textbf{Steric Clashes (Clashscore)}. 
    We calculated the steric clashscore~\cite{chen2010molprobity}, defined as the number of severe spatial overlaps per 1,000 heavy atoms. A steric clash is identified when the Euclidean distance between two non-bonded heavy atoms is less than the sum of their respective van der Waals (vdW) radii minus a tolerance margin of 0.4 Å. To strictly exclude covalently bonded neighbors, we parsed the molecular topology and only considered atom pairs separated by more than three bonds (i.e., shortest path > 3 in the chemical graph).
    \item \textbf{Backbone Geometry Deviations}.
    We measured the structural distortion of the protein backbone by computing the Mean Absolute Deviation (MAD) between the generated geometries and the standard ideal values (e.g., Engh \& Huber parameters~\cite{rossmann2001international}). Specifically, we analyzed the bond lengths for four primary backbone covalent bonds (N-C$\alpha$, C$\alpha$-C, C-N peptide bond, and C-O) and the bond angles for three primary backbone angles (N-C$\alpha$-C, C$\alpha$-C-N, and C-N-C$\alpha$).
    \item \textbf{Side-Chain Geometry Deviations}.
    Crucially, since our model generates full-atom coordinates including side chains, we explicitly evaluated the side-chain geometric quality. We utilized the standard amino acid templates from the PDB Chemical Component Dictionary (CCD) to construct ideal reference topologies. By mapping the generated side-chain heavy atoms to these standard templates, we systematically calculated the MAD for 93 distinct side-chain covalent bonds (e.g., CB-CG, CG-OD) and 134 distinct side-chain bond angles across all 20 standard amino acids.
\end{itemize}
The results are summarized in Table~\ref{tab:atomic}. The quantitative results demonstrate that our method achieves a notably lower steric clashscore compared to the strong baseline, indicating a more harmonious 3D spatial arrangement of atoms. Furthermore, our generated structures maintain highly precise local geometries, with both backbone and side-chain bond-length deviations strictly controlled to $\sim 0.02$ Å and bond-angle deviations to $\sim 2.0^\circ$. 

Additionally, we plot the side-chain torsion angle distributions in the generated samples for each canonical amino acid type in Figure~\ref{fig:chi_part_a} and \ref{fig:chi_part_b}, together with those obtained from the PDB and AFDB databases and the La-Proteina baseline as a reference. Our \ourmodel can produce structures that more closely resemble their natural occurrence, reflecting its better generative capability in capturing atomic conformations.
These minimal deviations fall well within the natural range of physical thermal fluctuations, strongly confirming the physical validity and high atomic fidelity of our fully atomic generation model.

\begin{figure}[p]
    \centering
    \begin{subfigure}{\textwidth}
        \centering
        \includegraphics[width=0.9\linewidth]{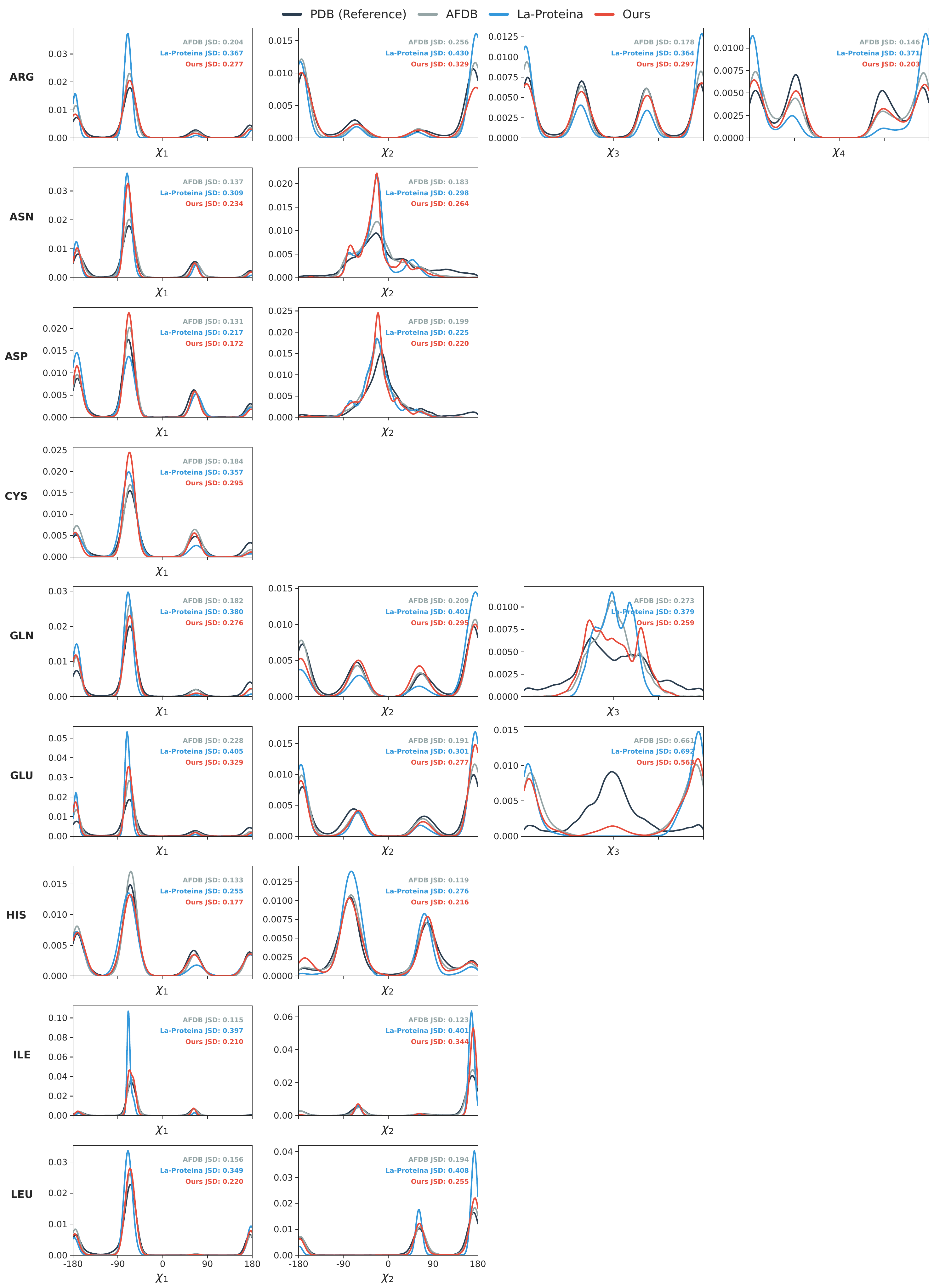}
        \caption{Distributions for residues ARG through HIS.}
        \label{fig:chi_part_a}
    \end{subfigure}
    \label{fig:chi_distribution_full}
\end{figure}

\begin{figure}[p]
    \ContinuedFloat
    \centering
    \begin{subfigure}{\textwidth}
        \centering
        \includegraphics[width=0.9\linewidth]{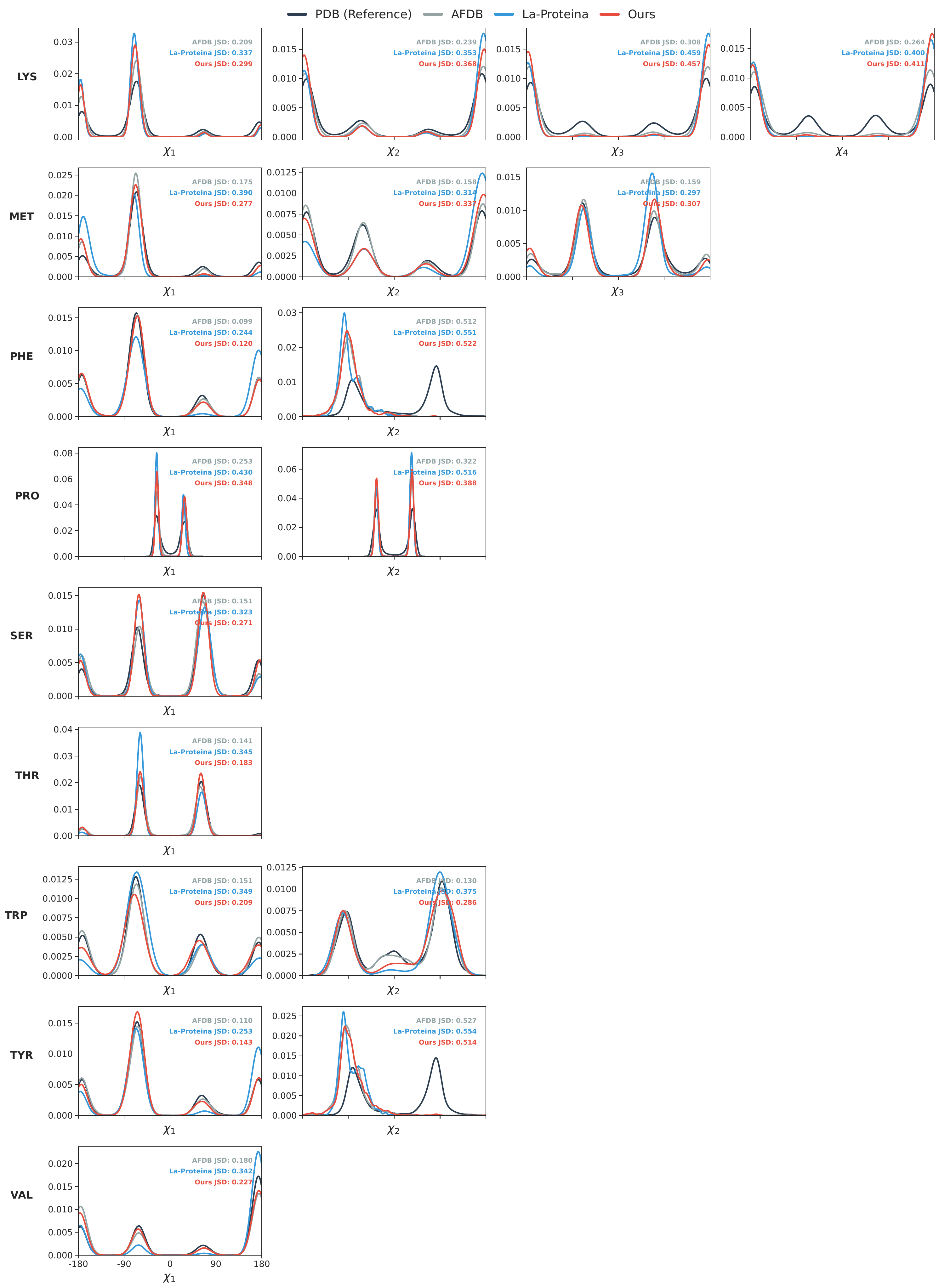}
        \caption{Distributions for residues ILE through VAL.}
        \label{fig:chi_part_b}
    \end{subfigure}
    \caption{Statistical distribution of side-chain dihedral angles.}
\end{figure}

To provide a more accurate understanding of generational diversity, we present additional analysis of the clusters obtained by \ourmodel and La-Proteina. In Table~\ref{tab:diversity}, we show that each designable structure cluster produced by our method is supported by substantially richer sequence diversity, with 4.19 sequence clusters per structure cluster on average (vs. 1.57 for La-Proteina) and much lower within-cluster pairwise sequence identity (0.410 vs. 0.872). This suggests that our method explores deeper sequence basins within the designable structural families it discovers.
Indeed, these results support the conclusion that the observed behavior reflects a genuine designability–diversity trade-off rather than a trivial artifact, as La-Proteina's structural diversity came with a pronounced drop in designability.
As future work, we will investigate whether broader structural coverage can be achieved more effectively through data-level interventions, such as training-data construction and sampling strategies, beyond sampling-time tuning alone.

\begin{table}[ht]
\centering
\caption{Analyses on the clusters generated by \ourmodel and La-Proteina.}\label{tab:diversity}
\begin{tabular}{@{}lrrr@{}}
\toprule
Method & \begin{tabular}[c]{@{}r@{}}Avg. Seq Clusters\\  / Str Cluster\end{tabular} & \begin{tabular}[c]{@{}r@{}}Within-Cluster \\ Pairwise Seq ID\end{tabular} & \begin{tabular}[c]{@{}r@{}}Top-5 Largest Str Clusters: \\ Avg. Seq Clusters\end{tabular} \\ \midrule
Ours & \textbf{4.19} & \textbf{0.410} & \textbf{18.6} \\
La-Proteina & 1.57 & 0.872 & 15.0 \\ \bottomrule
\end{tabular}
\end{table}

\subsection{Additional Results for Binder Design}
\begin{figure}[ht]
\centering
\includegraphics[width=.8\linewidth]{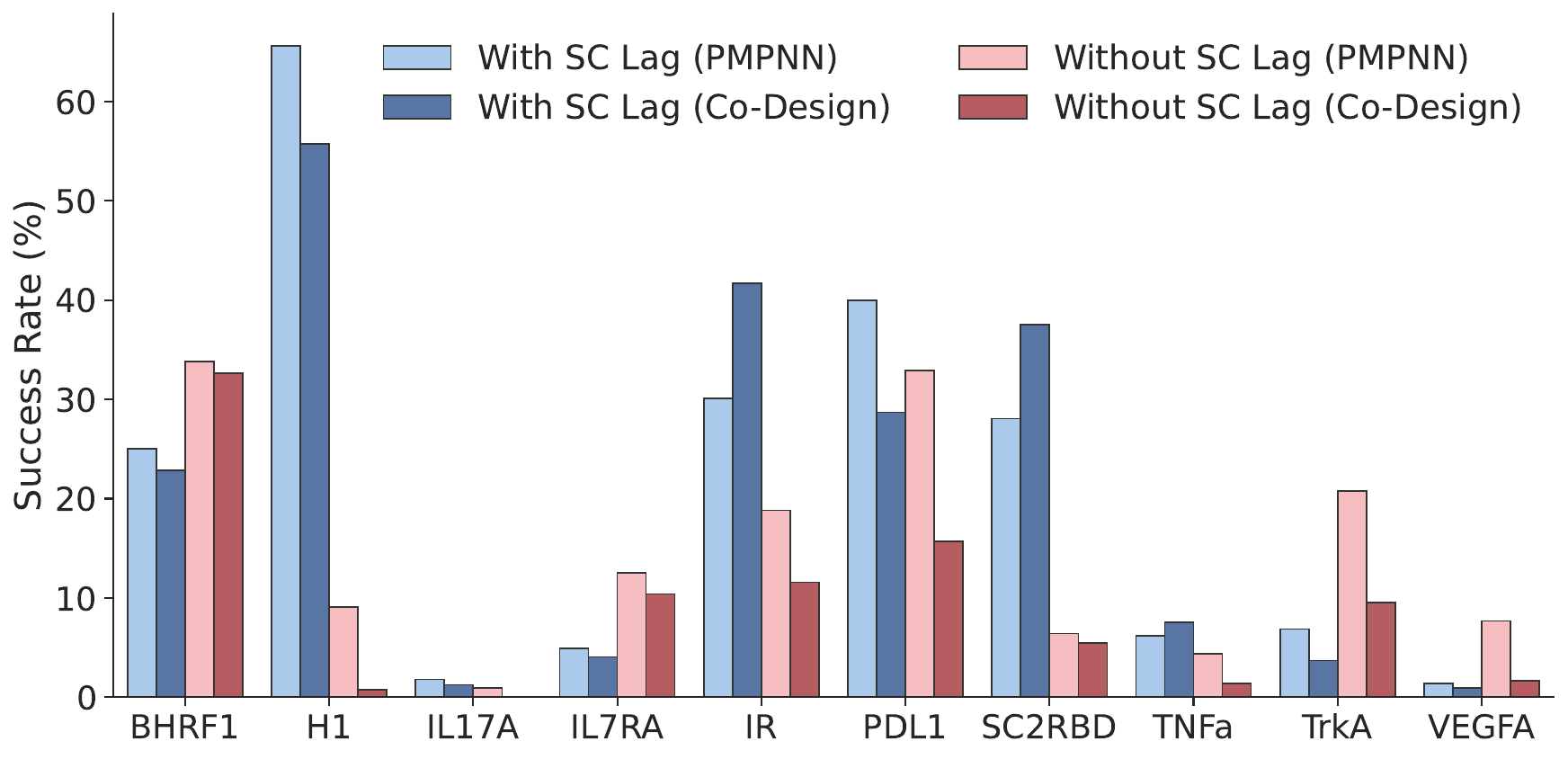}
\caption{Ablation study of side-chain lag training on binder design}
\label{fig:binder_ab}
\end{figure}

We benchmark the impact of side-chain lagging on the binder design generations in Figure~\ref{fig:binder_ab} across all ten targets. It can be seen that enabling the lag improves the success rates across 6 of 10 targets. Similar to the unconditional setup, we observe larger gains for co-designed sequences than for ProteinMPNN-redesigned ones.
This, together with the results from the unconditional benchmark, supports the effectiveness of our proposed side-chain lagging technique during sampling, mitigating the early commitment of side-chain positions that might otherwise lead to inconsistency between the continuous and discrete modalities.

\subsection{Additional Results for ncAA Modeling}\label{suppl:ncaa_result}
We provide a complete list of identified ncAA types (using exact element type and graph isomorphism matching as described in Appendix~\ref{suppl:ncaa_eval}), along with their scRMSDs relative to the reference CCD structures, in Table~\ref{fig:ncaa}. 
We also classified the generated ncAAs by their physicochemical properties in Figure~\ref{fig:ncaa_str}.
Within a limited number of structures generated and a low occurrence rate in the ncAA dataset of $\approx 1\%$, \ourmodel is still able to generate diverse identifiable ncAA structures, often at a high structure fidelity with scRMSD $<1$\AA.
The flexibility of \ourmodel's fully atomic design enables extensions that could incorporate ligands, DNA, and RNA into a unified protein design model.

\begin{table}[ht]
\centering
\caption{Complete list of identified ncAAs generated by \ourmodel with their scRMSD in the unconditional setup.}\label{tab:ncaa}
\begin{tabular}{@{}lrrrr@{}}
\toprule
CCD code & N samples & Mean scRMSD & Fraction scRMSD \textless 1\AA & Fraction scRMSD \textgreater 2\AA \\ \midrule
MLY & 103 & 0.854 & 0.951 & 0.000 \\
CEA & 74 & 0.842 & 0.716 & 0.000 \\
AA3 & 45 & 0.133 & 1.000 & 0.000 \\
2DO & 26 & 0.644 & 1.000 & 0.000 \\
CSD & 22 & 1.791 & 0.000 & 0.091 \\ \midrule
AA4 & 17 & 0.658 & 1.000 & 0.000 \\
CSS & 8 & 0.649 & 1.000 & 0.000 \\
5HP & 7 & 1.064 & 0.000 & 0.000 \\
XYC & 6 & 0.256 & 1.000 & 0.000 \\
OLT & 6 & 1.247 & 0.000 & 0.000 \\ \midrule
DDZ & 5 & 0.627 & 1.000 & 0.000 \\
MLZ & 5 & 0.715 & 1.000 & 0.000 \\
AS2 & 4 & 0.422 & 1.000 & 0.000 \\
CME & 4 & 1.122 & 0.250 & 0.000 \\
M3L & 3 & 2.782 & 0.000 & 1.000 \\ \midrule
MHO & 3 & 2.330 & 0.000 & 1.000 \\
DSE & 3 & 1.076 & 0.333 & 0.000 \\
FIO & 3 & 1.174 & 0.000 & 0.000 \\
02Y & 3 & 2.300 & 0.000 & 1.000 \\
4CY & 3 & 1.498 & 0.000 & 0.000 \\ \midrule
33X & 3 & 0.642 & 1.000 & 0.000 \\
HCS & 2 & 1.570 & 0.000 & 0.000 \\
OSE & 2 & 1.591 & 0.000 & 0.000 \\
0AZ & 2 & 1.173 & 0.000 & 0.000 \\
DGL & 2 & 0.226 & 1.000 & 0.000 \\ \midrule
CSU & 2 & 2.375 & 0.000 & 1.000 \\
A30 & 2 & 2.295 & 0.000 & 1.000 \\
HIX & 1 & 1.147 & 0.000 & 0.000 \\
KCX & 1 & 3.341 & 0.000 & 1.000 \\
FME & 1 & 1.445 & 0.000 & 0.000 \\ \midrule
0QL & 1 & 0.921 & 1.000 & 0.000 \\
BB6 & 1 & 1.442 & 0.000 & 0.000 \\
0A1 & 1 & 1.985 & 0.000 & 0.000 \\
ORD & 1 & 1.572 & 0.000 & 0.000 \\
DGN & 1 & 1.298 & 0.000 & 0.000 \\ \midrule
SCH & 1 & 0.721 & 1.000 & 0.000 \\
060 & 1 & 1.029 & 0.000 & 0.000 \\ \midrule
\textbf{Overall} & \textbf{375} & \textbf{0.872} & \textbf{0.736} & \textbf{0.043} \\ \bottomrule
\end{tabular}
\end{table}

\begin{figure}[ht]
\centering
\includegraphics[width=.95\linewidth]{figs/ncaa_str.jpg}
\caption{Generated ncAA structures classified by their physicochemical properties.}
\label{fig:ncaa_str}
\end{figure}


\end{document}